# Physically Constrained 3D Diffusion for Inverse Design of Fiber-reinforced Polymer Composite Materials

Pei Xu[1,#], Yunpeng Wu[2,#], Srikanth Pilla[3], Gang Li[2,*], and Feng Luo[1,*]

[1]School of Computing, AIM for Composites DOE-Energy Frontier Research Center, Clemson University, Clemson SC 29634

[2]Department of Mechanical Engineering, AIM for Composites DOE-Energy Frontier Research Center, Clemson University, Clemson SC, 29634

[3]Center for Composite Materials, AIM for Composites DOE-Energy Frontier Research Center, Department of Mechanical Engineering, University of Delaware, Newark DE, 19716

*To whom correspondence should be addressed:

Gang Li. Email: gli@clemson.edu

Feng Luo. Email: luofeng@clemson.edu

#These authors contributed equally to the manuscript as first authors.

**Abstract**

Designing fiber-reinforced polymer composites (FRPCs) with a tailored nonlinear stress-strain response can enable innovative applications across various industries. Currently, no efforts have achieved the inverse design of FRPCs that target the entire stress-strain curve. Here, we develop PC3D_Diffusion, a 3D spatial diffusion model designed for the inverse design of FRPCs. We generate 1.35 million FRPCs and calculate their stress-strain curves for training. Although the vanilla PC3D_Diffusion can generate visually appealing results, less than 10% of FRPCs generated by the vanilla model are collision-free, in which fibers do not intersect with each other. We then propose a loss-guided, learning-free approach to apply physical constraints during generation. As a result, PC3D_Diffusion can generate high-quality designs with tailored mechanical behaviors while guaranteeing to satisfy the physical constraints. PC3D_Diffusion advances FRPC inverse design and may facilitate the inverse design of other 3D materials, offering potential applications in industries reliant on materials with custom mechanical properties.

## Main

Fiber-reinforced polymer composite (FRPC) materials include short-fiber (or chopped fiber) reinforced composites and continuous fiber composites. These two kinds of composites have distinct physical characteristics and applications. Short-fiber reinforced composites are essential for industries where strength reinforcement is critical. They are more cost-effective and widely used in areas such as plastics, paper products, and certain textile fabrics. Continuous fiber composites are employed in applications demanding continuous reinforcement and high strength, often requiring specialized techniques. Designing FRPCs with a tailored nonlinear stress-strain response that allows for the customization of mechanical properties to meet specific performance criteria and enables innovative applications across various industries including aerospace [1], automotive [2], construction [3], health care [4], and others. For example, composite materials with tailored nonlinear responses can be used in applications requiring high energy absorption and dissipation capabilities [5, 6], such as protective gear, sports equipment, and automotive barriers. Another example involves soft composite materials designed to replicate the deformation response of human skin, including its subtle spatial variations across different areas of the body [7].

The inverse design of fiber-reinforced polymer composites with specific nonlinear stress-strain responses has unique challenges compared to other material design problems. First, the design space for these composites has an inherently 3D topological nature. Fiber-reinforced composites are inhomogeneous materials with complex microstructures. The spatial arrangement and orientation of fibers within the polymer matrix are critical to their mechanical properties. This complicates the design process by exponentially increasing the number of variables and the complexity of the relationships between them. Designing these composites requires a nuanced understanding of their 3D topological characteristics to achieve desired performance outcomes. Second, the non-linear stress-strain response of these composite materials originated by the material nonlinearity of the polymer matrix, influenced by their topological characteristics, complicates the prediction of their deformation behavior under different loading conditions and makes the inverse design a much more difficult task. Moreover, ensuring that the final design meets physical constraints, such as collision-free configurations and specific shapes (e.g., cylindrical), adds another layer of complexity. These constraints not only limit the design space but also necessitate adjustments to the 3D topology of the composite material to satisfy these requirements without compromising performance. Together, these factors make the inverse design of FRPCs with targeted nonlinear stress-strain responses a very challenging task. Traditional approaches that frame the material inverse design problem as an optimization task and solve it mathematically [8-10] are not adequate to thoroughly explore the design space of material microstructures for finding out the optimal design that satisfies not just a single load-deformation response, but the entire nonlinear stress-strain curve. To the best of our knowledge, no previous efforts have successfully achieved the inverse design of fiber-reinforced composites that target the entire stress-strain curve. This uncharted territory requires the development of methods capable of handling nonlinearity, navigating high-dimensional design spaces, and complying with strict physical constraints.

In recent years, the material inverse design field has made significant progress through the adoption of data-driven methods and deep learning techniques. Instead of seeking an explicit description to directly model the mathematical correlations between the design variables and the desired material performance, data-driven approaches aim to mine the relationships from large-scale data sets. Existing approaches utilize generative models like variational autoencoders (VAEs) [11, 12] and generative adversarial networks (GANs) [13, 14] to solve the inverse design problem of materials, or model the design problem as a sequential decision-making task and solve it via reinforcement learning [16-18]. While impressive results have been achieved in multiple material design tasks, such as molecular design [19], drug design [20], inorganic materials [21], thermoelectrics [22], structural topology optimization [23], metamaterials [15, 24], inverse design of FRPCs with targeted nonlinear stress-strain response has not yet been explored largely.

In the fields of computer vision and graphics, diffusion models [25-27] recently have become the state-of-the-art framework of generative models due to their capability to generate high-quality images. Diffusion models have also been applied to 3D spatial generation tasks, like protein backbone generation [28], targeted drug design [29], and rotation alignment [30]. However, existing vanilla methods do not fully consider the physical constraints during the generation and thus cannot guarantee the physical feasibility of the generated results. In this work, we present PC3D_Diffusion, a 3D diffusion model designed for the physically constrained inverse design of fiber-reinforced composites in 3D. To that end, we first programtically generates 1.35 million schmes of FRPCs and calculated their stress-strain curves via simulation. We adopt two distinct forms of diffusion models to process the position and orientation properties of fibers in 3D space. Additionally, we propose a loss-guided, learning-free approach to apply physical constraints during generation. PC3D_Diffusion allows precise control over the microstructural features of the composite, enabling the generation of materials with specific desired mechanical behaviors. Our evaluations showed that PC3D_Diffusion can generate high-quality design schemes whose stress-strain responses are close to the input one while guaranteeing the generation of physically constrained, collision-free fiber distributions.

## Results

### Generation of fiber-reinforced composites with stress-strain response curves

To train our diffusion model for inverse design, we generate a large-scale dataset of fiber-reinforced composites covering 52 fiber configurations (number of fibers per unit volume $n$ ranging from 10 to 50, fiber length $l \in \{30,50\}$ and fiber diameter $d \in \{4,6,8,10\}$ with unidirectional/heterogenous fiber orientations) in a 100mm×100mm×100mm representative volume element (RVE). We obtain their stress-strain response curves through finite element analysis (FEA) using ANSYS (Methods and Supplementary Notes). The FEA results show that the stress-strain curves could vary largely depending on the microstructure of the hyper-elastic fiber-reinforced polymer composites. Fig. S2 exhibits the ranges of the stress-strain curves of different fiber configurations.

For each configuration, we generate 26,000 samples and build a dataset with 1.35 million samples in total covering a wide range of stress-strain curves. This provides us an insight into the valid range of the stress-strain response of each fiber configuration. We take 25,000 samples from each fiber configuration as the training data and keep the remaining 1,000 samples for testing purposes.

### Inverse Design Process

Our inverse design process takes a target stress-strain curve as input and generates the microstructures of fiber-reinforced composites that satisfy the stress-strain curve (Fig. 1 a). We consider the inverse design process of FRPCs as a conditional generative task, where the position and orientation of each fiber are determined based on the given stress-strain curve ($s := [a_1, a_2, a_3] \in \mathbb{R}^3$) and the fiber configuration $c$ (including the fiber diameter ($d \in \mathbb{R}^1$), length ($l \in \mathbb{R}^1$), and fiber amount ($n$)). While $d$ and $l$ are given as extra conditional input to the model, the fiber amount condition $n$ is taken into account implicitly, which is decided by the model input dimension. Each fiber $i$ therefore has two attributes to be modeled: $\boldsymbol{p}_i \in \mathbb{R}^3$ is the position coordinate and $\boldsymbol{R}_i \in SO(2)$ is the orientation. The orientation $R_i$ is defined in the $SO(2)$ space, because the symmetry of the cylindric geometric model of fibers makes the rotation around the pole axis of each fiber itself inconsequential.

During generation, we first identify the proper fiber configurations for the expected stress-strain curve $s$ by matching the given stress-strain curve with the collected stress-strain response range of each configuration in our dataset (Methods). If the given stress-strain curve is out of the range of our dataset, we will guess a configuration based on those whose stress-strain response range is to the given stress-strain curve. Then, we feed the stress-strain curve and the chosen fiber configuration into our PC3D_Diffusion for FRPC generation (Fig. 1b). We perform conditional generation in a guidance-free way [31], where the condition is applied as a part of the input fed to the network for generation.

**3D Spatial Diffusion Model for Fiber Distribution Generation**

As a state-of-the-art technique of generative artificial intelligence, denoising diffusion models include a stochastic forward process and a reverse process. The forward process gradually adds noise through a Markov chain to convert a data point $x_0$ (e.g., the color of a pixel in an image) to a prior Gaussian distribution $\mathcal{N}(0, I)$ in $T$ steps. The reverse process takes a randomly drawn sample $x_t \in \mathcal{N}(0, I)$ as the initial value where $t = T$, and iteratively denoise the result from $x_t$ to $x_{t-1}$ until the final result $x_0$ is achieved. A neural network $\epsilon_\theta$ is employed to estimate the noise added to $x_0$ given $x_t$. Through the estimated noise, we can further estimate the $x_0$ based on given $x_t$, and then compute $q_\theta(x_{t-1}|x_t)$ from $q(x_{t-1}|x_t, x_0)$ by parameterizing $x_0$ using the estimated noise. In our case for FRPC generation, we add noise to the fiber positions $\{\boldsymbol{p}_i\}$and orientations $\{\boldsymbol{R}_i\}$ by randomly moving and rotating each fiber and performing denoising by rearranging the noised fiber distribution.

Due to the distinct characteristics of position and orientation, we use two ways to add noise to them. For the position property of fibers, we add Gaussian noise in a variance-preserving form [25] (Methods). The variance preserved form uses a scaling operation to eliminate the bias brought by the ground-truth value when adding noise and performing manipulation in the range of standard Gaussian distributions. However, it does not work well for properties that cannot support scaling operations. Therefore, for orientation, we adopt a variance exploding form [26] to apply rotation noise (Methods). While putting the two terms with respect to fiber positions and orientations together, we take a multi-objective learning technique [35] to automatically balance the learning process (Methods).

The reverse process starts with $t = T$. By drawing positions from normal Gaussian distributions and orientations from isotropic Gaussian distributions in SO space [34] as the initial fiber distribution $\{\boldsymbol{p}_T^i, \boldsymbol{R}_T^i\}_{i=1}^n$, inverse design is done through the reverse process of diffusion models, where $\{\boldsymbol{p}_T^i, \boldsymbol{R}_T^i\}_i$ is denoised into $\{\boldsymbol{p}_0^i, \boldsymbol{R}_0^i\}_i$ one step by step with $t$ decreases from $T$ to 0 (Methods). In this process, the fiber configuration (length $l$ and diameter $d$) and the coefficients of the given polynomial stress-strain curve are fed into the noise estimation network as the condition input. Fig. 1b shows the reverse process for fiber distribution generation.

**Noise Estimation Network for Diffusion Model**

The impact of one fiber on the overall stress-strain response is not only decided by the fiber's position and orientation in the RVE but also by interactions among multiple fibers. To better model the relations among multiple fibers, we employ a transformer architecture and utilize a stack of 32 transformer decoders as the backbone coupled with a graph attention network (GAT) [32] to synthesize the spatial representation of each fiber (Supplementary Notes). Overall, the network for noise estimation used by the diffusion model consists of (1) a graph attention network (GAT) to encode each fiber's 3D spatial representation, (2) a backbone of transformer decoder stack to perform noise estimation, and (3) a shared multilayer perceptron (MLP) to decode the transformer output to the predicted noise values added to the position and rotation properties for each fiber (Fig. 1a and Fig. S3). The total number of parameters including the shared GAT and MLP and the transformer backbone is around 1.5 million. The network is trained by minimizing two loss terms that measure the mean squared error of the predicted position and rotation noises respectively (Supplementary Notes).

Transformer architecture has been demonstrated capable of effectively capturing the relations between multiple parallel inputs [33]. When feeding the transformer network, we enhance the spatial representation of each fiber not only by taking into account the position and orientation of the fiber itself and also the relative positions and orientations of all neighbor fibers. To do so, we model fibers in an RVE as nodes in a graph with directed edges. The neighbor fibers are synthesized through the graph attention mechanism of GAT with learnable edge weights (Supplementary Notes). The spatial representation of each fiber is thus enhanced as $GAT\left(\boldsymbol{p}^i, \boldsymbol{R}^i, \{\boldsymbol{p}^j, \boldsymbol{R}^j\}_{j\neq i}\right)$. Projecting through MLP, the output of the neural network is the estimated noise added to $\{\boldsymbol{p}_0^i, \boldsymbol{R}_0^i\}$ to get $\{\boldsymbol{p}_t^i, \boldsymbol{R}_t^i\}$. An important characteristic of the transformer architecture is the input permutation invariant. Transformer decoders weigh each parallel input through an attention mechanism by taking into account the features of each input that are unrelated to the input order. For language processing tasks, this characteristic requires additional positional embeddings added to each input language token (words, characters, or phrases) to reflect the positional relations between the input tokens. However, this characteristic aligns with our setup, where the input order of fibers should not impact the output. Therefore, in our implementation, we directly apply the transformer decoder architecture to the enhanced fiber spatial representations without positional embeddings.

**Physical Constrained Generation of Fiber-reinforced Composites**

In actual composite materials, fibers do not penetrate or intersect with each other. The intersection of the fibers is referred to as "collision" in this work. Although the vanilla 3D spatial diffusion model can generate visually appealing results, its reverse process relies on random sampling, making it difficult to directly generate a collision-free fiber distribution. As shown in Fig. 2b, even in the most simple test case during experiments with the configuration of $n = 10, d = 10, l = 50$ and heterogeneous orientations, there are less than 10% of generated composites are collision-free. For heterogeneous fiber orientations, the percentage of collision-free generation is even lower than 4% for ($n = 30, d = 10, l = 50$) and 0.2% for ($n = 30, d = 10, l = \infty$). On the other hand, we observed that the collisions between fibers are typically subtle in the generated (Fig. 2a). This implies that we can correct the generated fiber distributions to achieve collision-free results by slightly moving or rotating the fibers in the generated FRPCs.

Here, we introduce a guidance loss term during the reverse processing to apply physical constraints and enforce collision-free generation. We use a differentiable constraint loss function measuring the distance between each pair of fibers with a boundary constraint:

$$\mathcal{L}_{\text{cons}}\left(\{\boldsymbol{p}_t^i,\boldsymbol{R}_t^i\}_i\right)=\frac{1}{n}\sum_{ij} max\left\{0,1-\frac{DIST\left(SHRINK(\boldsymbol{p}_t^i),\boldsymbol{R}_t^i;SHRINK\left(\boldsymbol{p}_t^j\right),\boldsymbol{R}_t^j\right)}{d+\varepsilon}\right\} \quad (1)$$

where $d$ is the diameter of fibers, ε is the minimal gap allowed between two fibers, $DIST(\cdot;\cdot)$ is a differentiable distance function and $SHRINK(\cdot)$ function shrinks fiber length to ensure the boundary constraints (Supplementary Notes). $\mathcal{L}_{\text{cons}}=0$ when there is no collision, i.e. $DIST\left(p_t^i,R_t^i;p_t^j,R_t^j\right)\geq d+\varepsilon$ for all fiber pairs. At each step $t$ during the reverse process, we minimize the loss guidance term by

$$\{\boldsymbol{p}_t^i,\boldsymbol{R}_t^i\}_i \leftarrow \{\boldsymbol{p}_t^i,\boldsymbol{R}_t^i\}_i-\alpha\nabla_{\{\boldsymbol{p}_t^i,\boldsymbol{R}_t^i\}_i}\mathcal{L}_{\text{cons}}\left(\{\boldsymbol{p}_t^i,\boldsymbol{R}_t^i\}_i\right). \quad (2)$$

Applying $L_{\text{cons}}$ to update fiber positions and orientations is an optimization process of gradient descent without needing any learning. In our implementation of the reverse process, we update $\{\boldsymbol{p}_t^i,\boldsymbol{R}_t^i\}$ once at each time step $t$ and keep performing updates until $\mathcal{L}_{\text{cons}}\left(\{\boldsymbol{p}_t^i,\boldsymbol{R}_t^i\}_i\right)=0$ at the last step from $t=1$ to 0. Since our 3D spatial diffusion model can directly provide results where fibers would not largely collide with others, typically we can achieve collision-free results within 10 iterations at the last step, while the whole reverse process takes around 8 seconds given $T=500$ in our implementation.

Theoretically, when generating unidirectional fiber distributions, we can add an extra loss guidance term measuring the orientation divergence of fibers. However, during experiments, we found that the 3D spatial diffusion model trained with our dataset of unidirectional fiber orientations can directly provide unified orientations with very small errors (see the next section for quantitative results). Therefore, for fiber configurations with unidirectional orientations, we perform updating only on fiber positions $\{\boldsymbol{p}_t^i\}_i$ in the reverse process of fiber distribution generation.

By applying the loss guidance, our approach can generate collision-free results even in very dense scenarios. From the examples shown in Fig 2 and Fig S9, the overall distributions of fibers with and without the loss guidance are almost identical. The introduction of loss guidance only slightly changes the positions and/or orientations of fibers without significant modification to the overall distributions of fibers. Although the guided loss itself does not consider the stress-strain response difference when modifying the fiber distribution, the stress-strain response is still guaranteed by the following generation steps. At the last step of generation, the guided loss is applied iteratively without further revision. However, it is observed that the generated results are already near collision-free, as we achieve completely collision-free results typically with less than 10 iterations during our testing.

### Generation of Fiber-reinforced Composites using Conditions within the Range of Training Dataset

We first demonstrate our model's capability to generate FRPCs for stress-strain curves and configurations within the range covered in our training dataset, as shown in Fig. S2. For each configuration, we use a test set of 1000 samples with stress-strain curves not presented in the training dataset. For a given stress-strain curve, our system provides multiple candidate fiber configurations. The stochastic nature of the reverse process of data generation using diffusion models makes it very easy for our approach to generate multiple results simultaneously. Here, in Fig. 3, for each candidate fiber configuration ($n$, $l$, $d$, heterogeneous or unidirectional orientations), we visualize six distinct material designs generated by our approach with volume fractions varying from 10% to 18%. We verify the generated results by FEA. All generated results of the five fiber configurations have stress-strain responses very close to the input curve. More examples of generation are shown in Fig. S4-8. It is shown that our method can generate high-quality results with diverse fiber configurations and volume fractions (from 1.70% to 19.56%), all producing stress-strain curves that closely match the input data. In cases of unidirectional fibers, the variation in fiber orientation is less than 1 degree in all the test cases without applying loss guidance during generation. Moreover, the generated results are consistent with the underlying mechanics of composite materials. For short fibers with heterogeneous orientations, given the same fiber length and diameters, our model generates more fibers, leading to an increased volume fraction, in response to an increase in the required composite stiffness (i.e., the slope of the input stress-strain curve), as shown in Fig. S4-6. In cases of unidirectional fibers, our model increases the number of fibers along the loading direction to produce a stiffer stress-strain curve (Fig. S7-8). Such behavior demonstrates that the physical principles of the composite mechanics have been learned and understood by our model.

Next, we evaluate our method quantitatively using the testing dataset (Methods). In practice, after generating multiple fiber distributions, we can easily find out the best after running verification using FEA. Therefore, we first measure the upper-bound performance of our methods. We select the best-of-10 generated results for each testing sample based on the relative error area $e_A$ among every 10 generated results. As shown in Table 1, the average mean-absolute-errors (MAEs) between expected curve coefficients ($a_1$, $a_2$ and $a_3$) and the ones of generated results ($\hat{a}_1$, $\hat{a}_2$ and $\hat{a}_3$) are all less than 0.5%, and their $e_A$ are less than 0.2%. For composites with unidirectional fibers, the fiber orientation divergence has a standard deviation of less than 0.08°.

We further evaluate the average performance of 10 generated composites (Table S4). For generated composites with heterogeneous orientations, the average MAEs are all less than 1% and $e_A$ are less than 0.45%. The generated composites with unidirectional fibers have higher MAEs (≤2.42%) and $e_A$ (≤0.96%). The maximal error of 2.42% occurs at $\hat{a}_3$ for the configuration of $n = 30, l = \infty, d = 10$ with unidirectional fiber orientations. The orientation divergence of composites with unidirectional fibers only has a standard deviation of less than 0.31°. We can, thereby, simply regard that fibers all have the same orientations in the generated results. Note that the impact of the error of the 3rd-order coefficient $a_3$ is much smaller than that of the 1st-order coefficient $a_1$. As shown in the Table 1 and S4, $e_A$ is identical with $e_{a_1}$ for almost all configurations, which means that the error between the generated and the target stress-strain curves mostly comes from the error of $e_{a_1}$. Overall, the small value of $e_A$ indicates that our model can consistently provide composite designs with stress-strain curves that closely match the expected ones.

**Generation using Out-of-the-Range Conditions**

To further evaluate the generalization capability of the PC3D_Diffusion model, we examine its performance on stress-strain curves and configurations out of the range of the training datasets. First, we consider configurations that are outside of the training data for a stress-strain curve that falls within the range of the training dataset. As shown in Fig. 4a-4d, all four generated fiber distributions with configurations not covered by the training set give stress-strain curves that are very close to the expected ones.

Next, we consider a more challenging task where both the configuration and stress-strain curves fall outside the range of our training data. We select a stress-strain curve that is below the envelope of stress-strain curves for heterogeneous fiber orientations, and another one below the envelope of stress-strain curves for unidirectional fibers. For both stress-strain curves, we select the configuration $n = 30, l = 25, d = 2$ where $l$ and $d$ are out of range of the training data. As shown in Fig. 4e-4f, the composites generated by the PC3D_Diffusion model can produce stress-strain curves that closely match the expected one, while the given curves fall out of the overall curve ranges of the training data. These examples show the generalization capability of our model. It is important to note that this generalization capability allows for the discovery of new composite material designs with superior material responses that have not been seen before.

**Model Analysis**

We have examined how transformer architecture and fiber representation impact the performance of our model. We use the configuration of $n = 30, l = 50, d = 10$ for the study, and take the same approach as our quantitative evaluation to perform model analysis. For each testing sample, we generate 10 results and use the average performance to evaluate the models. As shown in Table S5, the model performance increases consistently as the model scale, i.e. the number of attention heads and the number of decoder layers, increases. To evaluate the performance of the model without using GAT for fiber spatial representation learning, we simply pass each fiber state $[\boldsymbol{p}^i, \boldsymbol{R}^i \in \mathbb{R}^5]$ through a shared network with two fully connected layers to get a 512-dimension vector as the embedding of each fiber fed into the transformer network. The models with or without GAT have a similar size (number of parameters). However, the GAT can reduce the average errors $e_{a_1}$, $e_{a_2}$, $e_{a_3}$and $e_A$ by around 25% (Table S5).

## Discussion

Designing fiber-reinforced composites in 3D space is challenging for traditional optimization methods due to the large design space of material microstructure. The challenge would be further amplified when requiring the design to satisfy the entire nonlinear stress-strain curve. In this work, we regard the inverse design problem of fiber-reinforced composites as a generative task to decide the distribution of fiber positions and orientations. Our 3D diffusion model considers the distinct characteristics of position and orientation properties and is particularly suitable for 3D fiber-reinforced composites under a non-linear mechanical deformation setting. By collecting 1.3 million samples for training, our data-driven model can accurately capture and reproduce the entire non-linear stress-strain response under large deformation, which is difficult, if not impossible, to achieve with traditional optimization methods.

While the composite microstructure configuration for a given stress-strain response is inherently not unique, the stochastic nature of the diffusion model allows us to easily and parallelly generate various valid composite designs. This feature offers composite engineers multiple options to consider, along with other factors like manufacturability and cost. In comparison, traditional optimization methods typically stop searching when one design is found. As demonstrated by the results, the generalization capability of the 3D diffusion model enables the generation of compo-site designs to previously unseen stress-strain curves, facilitating the discovery of new composite materials exhibiting unprecedented properties.

The current study is designed to generate fiber-reinforced polymer composites for given stress-strain curves. Our approach is generally applicable to any spatial generation task. It can be easily expanded to other types of composites such as particulate, flake, or laminar composites, and also target other material properties such as viscoelastic, damage, and/or fatigue properties, albeit with additional training data, probably a larger network, and more training time.

Designs generated by our vanilla 3D spatial diffusion model have a low percentage of collision-free results since there is no mechanism in the vanilla 3D diffusion model to enforce physical constraints. To guarantee the generation of physically allowed fiber distributions, we propose distance-guided generation to avoid collision between fibers. This loss guidance approach can be further extended by introducing other differentiable losses to apply additional physical constraints, e.g., fiber packing density limits imposed by manufacturing processes.

Although our model shows high-quality results in generating fiber distribution in 3D space, it has limitations. The generated results are not always realizable due to the difficulty of precisely controlling individual fiber's position and orientation in the manufacturing process. As a direction for further research, we are exploring approaches to take into account manufacturability in generating fiber distributions. Experimental validations will be carried out for the generated "manufacturing-aware" composite designs. For an arbitrarily given stress-strain curve, there may not be a corresponding composite design solution. In such cases, our current model may generate designs that best approximate the target curve. However, it is desirable to determine a priori the feasibility of an input stress-strain curve to ensure the existence of a material design solution. To this end, an auxiliary mechanism may be necessary to help decide what is a feasible stress-strain curve. In addition to these limitations, it should be noted that the training data used by the 3D diffusion model is produced by the deterministic finite element analysis (FEA) model. In practice, composite materials often display small variability in properties due to variations in the matrix properties, the size, and geometry of the fibers, as well as the conditions under which they are manufactured. To capture such uncertainties, it is necessary to employ a probabilistic approach (e.g. stochastic FEA models) to generate training data, which is beyond the scope of this work.

# Methods

Here, we provide details of data generation and the implementation of PC3D_Diffusion. We refer to the supplementary materials for more details and mathematical inference of our implementations.

## Database Generation

We generate 52 combinations of fiber configurations with the number of fibers in a 100mm×100mm×100mm representative volume element (RVE) in the range between 10 and 50 and diameters of 4, 6, 8, and 10 in the unit of millimeters. We consider short fibers with lengths of 30mm and 50mm and long fibers completely penetrating the RVE.

We consider two orientation configurations: heterogeneous orientations where all fibers have random and independent orientations, and unidirectional orientations where all fibers in an RVE have the same orientation. Given a fiber configuration, we generate samples by randomly placing fibers into the RVE one by one with a collision check to avoid fibers intersecting or overlapping with others. Collision-free placement is the prerequisite for finite element analysis (FEA) and the physical constraint that must be met during inverse design.

All fibers are modeled as homogeneous cylinders and considered linear materials, while the polymer matrix is modeled as hyperelastic materials for large nonlinear elastic deformation. Each fiber cylinder is discretized as an octagonal prism to generate tetrahedral meshes for FEA with external parts outside the RVE cut off along the six sides of the RVE cube.

Stress-strain responses are obtained through simulation by applying displacements of 10 mm, 20 mm, and 30 mm along the x-direction of the RVE, which corresponds to 10%, 20%, and 30% strains respectively. The stress-strain curve is fitted as a cubic function:

$$\sigma(\eta) = a_1\eta + a_2\eta^2 + a_3\eta^3$$

where $\sigma$ is the nominal stress, $\eta$ is the strain, and $a_1$, $a_2$ and $a_3$ are coefficients fitted based on the given simulated stress results $\sigma(10\%)$, $\sigma(20\%)$, and $\sigma(30\%)$. Fig. S2 shows the ranges of the stress-strain curves of different fiber configurations.

More details of the data generation process are described in the Supplementary Notes.

## Diffusion on Fiber Positions

For the position property of fibers, we add Gaussian noise in a variance-preserving form [25]:

$$q(\boldsymbol{p}_t^i|\boldsymbol{p}_{t-1}^i) = \mathcal{N}(\boldsymbol{p}_t^i; \sqrt{1-\beta_t}\boldsymbol{p}_{t-1}^i, \beta_t\boldsymbol{I}) \tag{3}$$

given the variance schedule $\{\beta_t \in (0,1)\}_{t=1}^T$ where $\beta_t$ increases with $t$. Through the Markov chain, we have

$$q(\boldsymbol{p}_t^i|\boldsymbol{p}_0^i) = \mathcal{N}\left(\boldsymbol{p}_t^i; \sqrt{\overline{\alpha}_t}\boldsymbol{p}_0^i, (1-\overline{\alpha}_t)\boldsymbol{I}\right) \tag{4}$$

where $\overline{\alpha} = \prod_{\tau \le t} \alpha_\tau$, and $\alpha_\tau = 1 - \beta_\tau$ . As $t$ increases to $T$, $1 - \overline{\alpha}_t$ also increases and reaches near 1 when $t = T$. Correspondingly, $\sqrt{\overline{\alpha}_t}$ decreases while $t$ increases, and reaches near 0 when $t = T$. This setup approximately leads to $q(\boldsymbol{p}_T^i) \sim \mathcal{N}(\boldsymbol{0}, \boldsymbol{I})$, where the bias of $\boldsymbol{p}_0^i$ is roughly scaled down to zero and the variance is preserved with the noise $\epsilon$. During the reverse process for generation, given $\boldsymbol{p}_{\epsilon_t}^i$ as the noise added to $\boldsymbol{p}_0^i$ at the step $t$, we can draw $\boldsymbol{p}_{t-1}^i$ from $q(\boldsymbol{p}_{t-1}^i|\boldsymbol{p}_t^i)$ by replacing $\boldsymbol{p}_0^i$ with $\boldsymbol{p}_0^i = \left(\boldsymbol{p}_t^i - \sqrt{1-\overline{\alpha}_t}\boldsymbol{p}_{\epsilon_t}^i\right)/\sqrt{\overline{\alpha}_t}$ in $q(\boldsymbol{p}_{t-1}^i|\boldsymbol{p}_t^i, \boldsymbol{p}_0^i)$.

Following DDPM, we train the position diffusion using the loss function:

$$\mathcal{L}_{\boldsymbol{p}} = \mathbb{E}\left[\frac{1}{n}\sum_i \left\|\boldsymbol{p}_{\epsilon_t}^i - \epsilon_{\boldsymbol{p}}\left(\boldsymbol{p}_t^i, \boldsymbol{R}_t^i, \{\boldsymbol{p}_t^i, \boldsymbol{R}_t^i\}_{j\neq i}, c, t|\theta\right)\right\|^2\right] \quad (5)$$

where $\boldsymbol{p}_t^2 = \sqrt{\overline{\alpha}_t}\boldsymbol{p}_0^i + \sqrt{1-\overline{\alpha}_t}\boldsymbol{p}_{\epsilon_t}^i$, $\boldsymbol{p}_{\epsilon_t}^i \sim \mathcal{N}(\boldsymbol{0}, \boldsymbol{I}) \in \mathbb{R}^3$, the original data $\boldsymbol{p}_0^i$ is normalized such that the cubic RVE measuring 100 mm×100 mm×100 mm is projected to the range of $[-1, 1]$ along all three dimensions, and $\epsilon_{\boldsymbol{p}}$ is the noise prediction network optimized with parameter $\theta$.

**Diffusion on Fiber Orientation**

We approximate Gaussian-like sampling for rotation based on isotropic Gaussian distribution in SO space [34]. The isotropic Gaussian parameterizes a rotation in an axis-angle form, where the axis $\boldsymbol{u}$ is sampled uniformly and rotation angle $\omega$ has a probabilistic density function defined in a discrete form:

$$f_{IGSO}(\omega|\sigma^2) = \frac{1-cos\omega}{\pi}\sum_{l=0}^{L}(2l+1)\,exp(-l(l+1)\sigma^2)\frac{sin((l+1/2)\omega)}{sin(\omega/2)} \quad (6)$$

where $\omega \in [0, \pi]$ and $\sigma^2$ is the variance. Given that fibers are considered as homogeneous cylinders, the effective rotation range is only $[-\pi/2, \pi/2]$, we further scale $\omega$ by $1/2$ after the computation of probabilistic density. The distribution of the rotation angle $\omega$, thereby, is still Gaussian-like but in the range of $[0, \pi/2]$ after scaling. The direction of the rotation angle is decided by the rotation axis and we thus can always assume $\omega \geq 0$. For each fiber $i$, a rotation noise $\boldsymbol{R}_{\epsilon_t}^i$ is obtained through the rotation axis $\boldsymbol{u}_t^i = \widetilde{\boldsymbol{u}}_t^i/\|\widetilde{\boldsymbol{u}}_t^i\|$, where $\widetilde{\boldsymbol{u}}_t^i \sim \mathcal{N}(\boldsymbol{0}, \boldsymbol{I}) \in \mathbb{R}^2$ is the rotation direction vector that we ignore the component of the rotation around the fiber itself and the rotation angle $\omega_t^i \sim IGSO(\sigma_t^2)$. The rotation with added noise is defined as

$$\boldsymbol{R}_t^i = \boldsymbol{R}_0^i\left(\boldsymbol{R}_{\epsilon_t}^i\right)^T . \quad (7)$$

The vanilla variance exploding form takes $x_t \sim \mathcal{N}(x_t; x_0, \sigma_t^2 I)$. It requires $\sigma_t$ to be quite large in order to suppress the bias brought by $x_0$. Due to the "wrappable" nature of rotation angles, when we apply the formula for rotations by replacing the add operation with rotation multiplication, the bias of $\boldsymbol{R}_0^i$ can be eliminated as long as the cumulative distribution function of $IGSO(\sigma_t^2)$ is approximately linear in the range of $[0, 2\pi]$. We adopt a quadratic schedule with $\sigma_t \in (0.05, 5)$. This results in a near-linear cumulative distribution function of $IGSO(5^2)$ given $\sigma_t = 5$ when $t = T$. We train the neural network to predict the noise rotation in the form of exponential mapping:

$$\mathcal{L}_{\boldsymbol{R}} = E\left[\frac{1}{n}\sum_i \left\|\omega_t^i\boldsymbol{u}_t^i - \epsilon_{\boldsymbol{R}}\left(\boldsymbol{p}_t^i, \boldsymbol{R}_t^i, \{\boldsymbol{p}_t^i, \boldsymbol{R}_t^i\}_{j\neq i}, c, t|\theta\right)\right\|^2\right] . \quad (8)$$

**Combining the Diffusion on Fiber Position and Orientation**

To balance the learning of the two objectives, we perform multi-objective learning by introducing two additional learnable variables $w_{\boldsymbol{p}}$ and $w_{\boldsymbol{R}}$ to automatically adjust the weights of the two objectives [35]:

$$\mathcal{L} = \frac{1}{w_{\boldsymbol{p}}^2}\mathcal{L}_{\boldsymbol{p}} + \frac{1}{w_{\boldsymbol{R}}^2}\mathcal{L}_{\boldsymbol{R}} + 2\log w_{\boldsymbol{p}} w_{\boldsymbol{R}} \ . \quad (9)$$

### Reverse Process for Composite Generation

We employ the vanilla denoising diffusion probabilistic model to denoise the position property of fibers. The reverse process of fiber position is done iteratively by

$$\begin{cases} \boldsymbol{p}_{t-1}^{i} = \epsilon_{\boldsymbol{p}}\left(\boldsymbol{p}_{t}^{i}, \boldsymbol{R}_{t}^{i}, \{\boldsymbol{p}_{t}^{i}, \boldsymbol{R}_{t}^{i}\}_{j\neq i}, c, t|\theta\right) & \text{for } t = 1 \\ \boldsymbol{p}_{t-1}^{i} \sim \mathcal{N}\left(\frac{1}{\sqrt{\overline{\alpha}_t}}\left(\boldsymbol{p}_{t}^{i} - \frac{\beta_t}{\sqrt{1-\overline{\alpha}_t}}\epsilon_{\boldsymbol{p}}\left(\boldsymbol{p}_{t}^{i}, \boldsymbol{R}_{t}^{i}, \{\boldsymbol{p}_{t}^{i}, \boldsymbol{R}_{t}^{i}\}_{j\neq i}, c, t|\theta\right)\right), \tilde{\beta}_t I\right) & \text{otherwise} \end{cases} \tag{10}$$

where $t \in [1, T]$, $\boldsymbol{p}_{T}^{i} \sim \mathcal{N}(\boldsymbol{0}, \boldsymbol{I})$ is sampled randomly and $\tilde{\beta}_t = (1-\overline{\alpha}_{t-1})\beta_t/(1-\overline{\alpha}_t)$. We employ a quadratic variance schedule in the range of $[\beta_1, \beta_T]$: $\beta_t = \left(\frac{T-(t-1)}{T}\sqrt{\beta_1} + \frac{t}{T}\sqrt{\beta_T}\right)^2$.

For the reverse of fiber orientation, we have a diffusion process defined in a variance exploding form [26] without suppressing the groud-truth value during diffusion:

$$\boldsymbol{R}_{t-1}^{i} = \boldsymbol{R}_{t}^{i}\boldsymbol{R}_{\epsilon_{t\to t-1}}^{i} \tag{11}$$

where $\boldsymbol{R}_{\epsilon_{t\to t-1}}^{i}$ is the inverse of the noise rotation added to obtain $\boldsymbol{R}_{t}^{i}$ from $\boldsymbol{R}_{t-1}^{i}$. We use a score-based approach [27] to perform the reverse process of generation and obtain the estimated noise rotation $\widehat{\boldsymbol{R}}_{\epsilon_{t\to t-1}}^{i}$ through the exponential map representation $\omega_{\epsilon_{t\to t-1}}^{i}\boldsymbol{u}_{\epsilon_{t\to t-1}}^{i}$ by

$$\begin{cases} \omega_{\epsilon_{t\to t-1}}^{i}\boldsymbol{u}_{\epsilon_{t\to t-1}}^{i} = \widehat{\omega}_{t}^{i}\widehat{\boldsymbol{u}}_{t}^{i} & \text{for } t = 1 \\ \omega_{\epsilon_{t\to t-1}}^{i}\boldsymbol{u}_{\epsilon_{t\to t-1}}^{i} \sim \mathcal{N}\left(-\widehat{\boldsymbol{u}}_{t}^{i} g(\bar{t})^2 \mathrm{d}\bar{t} \nabla log f_{IGSO}(\widehat{\omega}_{t}^{i}|\sigma_{\bar{t}}^{2}), g(\bar{t})^2 \mathrm{d}\bar{t} I\right) & \text{otherwise} \end{cases} \tag{12}$$

where $\widehat{\omega}_{t}^{i}\widehat{\boldsymbol{u}}_{t}^{i} = \epsilon_{\boldsymbol{R}}\left(\boldsymbol{p}_{t}^{i}, \boldsymbol{R}_{t}^{i}, \{\boldsymbol{p}_{t}^{i}, \boldsymbol{R}_{t}^{i}\}_{j\neq i}, c, t\right)$ is the estimated noise $\widehat{\boldsymbol{R}}_{\epsilon_{t\to t-1}}^{i}$ provided by the noise estimation network and written in the exponential map with the magnitude $\widehat{\omega}_{t}^{i}$ and the direction vector $\widehat{\boldsymbol{u}}_{t}^{i}$, $\bar{t} = t/T$ is the normalized time step, and $f_{IGSO}(\cdot\,|\sigma^2)$ is the probabilistic density function of the distribution $IGSO(\sigma_{t}^{2})$. We adopt a quadratic variance schedule: $\sigma_{\bar{t}} = \sigma_0 + b_0\bar{t} + (\sigma_1 - \sigma_0 - b_0)\bar{t}^2$. Based on the inference from [27], $g(\bar{t})^2$ is obtained by the derivative of $\sigma_{\bar{t}}^{2}$ with respect to $\bar{t}$, and thus we have $g(\bar{t})^2 = \mathrm{d}\sigma_{\bar{t}}^{2}/\mathrm{d}\bar{t} = 2\sigma_t\left(b_0 + 2\bar{t}(\sigma_1 - \sigma_0 - b_0)\right)$.

### Inverse Designing Process

We consider the inverse design process of FRPCs as a conditional generative task of deciding the position and orientation of each fiber based on the coefficients of the given stress-strain curve and fiber configuration. We employ the denoising diffusion model as the fundamental framework for our generative inverse design. In the diffusion process of model training, we add noise to fiber positions and orientations based on the randomly sampled, discrete-time variable $t$ drawn from the range between 1 and $T$ uniformly. A neural network is trained in this process to estimate the added noise given the noised fiber positions and orientations (Equations 5, 8 and 9).

During inference time for inverse design, given an expected stress-strain curve, we first choose the target fiber configuration based on the coefficients ($a_1$, $a_2$, and $a_3$ in Equation 1) of the target stress-strain curve. This process is done by finding out the fiber configurations whose valid stress-strain curve ranges can cover the provided curve. The fiber length ($l$) and diameter ($d$) of the found target fiber configuration will be given to the model as a condition vector. The fiber amount ($n$) is implicitly imposed by proposing a set of $n$ fiber positions and orientations $\{\boldsymbol{p}_t^i, \boldsymbol{R}_t^i\}_{i=1}^n$ as the input to the proposed transformer-based diffusion model. Then, in the reverse process for generation, we start by randomly drawing a set of fiber positions $\boldsymbol{p}_T^i$ and orientations $\boldsymbol{R}_T^i$, and utilize the trained noise estimation neural network to iteratively denoise the drawn results based on Equations 10 and 12. Additionally, we introduce a loss-based, learning-free guidance (Equation 1) during the reverse process to apply physical constraints on the generated results and to ensure the generated fiber distributions free of collisions. The whole reverse process is elaborated in Algorithm 2 in the supplementary materials. We also refer to the supplementary materials for the details of the inference protocol for inverse design.

### Quantitative Evaluation

To evaluate our approach quantitatively, we test our model using the testing dataset containing 1,000 samples for each fiber configuration. This leads to a test set of 52,000 samples. We take the coefficients of the fitted stress-strain curves of the samples in the test set as the condition input to generate fiber distributions and verify the generated results using FEA. We compare the given, expected curve coefficients ($a_1$, $a_2$ and $a_3$) to the ones obtained by simulation on the generated results ($\hat{a}_1$, $\hat{a}_2$ and $\hat{a}_3$) through relative mean absolute error (MAE):

$$e_{a_i} = \left|\frac{a_i - \hat{a}_i}{a_i}\right| \times 100\% \tag{13}$$

where $i = 1,2, and$ 3 given the stress-strain curve in the form of Equation 1.

Since our stress-strain curves are modeled as a cubic function with three orders, the error contribution from the coefficient of each order has a different basis. To compare the expected, input curve and the generated curve more intuitively, we compute the relative error area between the two curves:

$$e_A = \frac{\int_{0.1}^{0.3} \left|\sum_{i=1}^{3} a_i \eta^i - \sum_{i=1}^{3} \hat{a}_i \eta^i\right| d\eta}{\int_{0.1}^{0.3} \sum_{i=1}^{3} a_i \eta^i d\eta} \tag{14}$$

given that the valid range of the strain $\eta$ is $[0.1, 0.3]$ during our simulation. The smaller $e_A$ is, the closer the expected and generated curves are.

### Data Availability

The whole database including the training and validation data is available at https://drive.google.com/drive/folders/1ezahBsw5ogX1JilRAmnFWFdTE1KVkiGz.

### Code Availability

The code implementation for our diffusion model for fiber distribution generation is available at https://github.com/TheLuoFengLab/PC3D_Diffusion.

## Author contributions

F.L., G.L. and S.P. conceived and designed this project. Y.W. generated FRPC data and calculated the properties. P.X. developed and implemented the PC3D_diffusion model. P.X. and Y.W. performed the evaluation. P.X., S.P., G.L., and F.L. wrote the manuscript. All authors have read and approved the final version of this manuscript.

## Acknowledgments

This work was supported as part of the AIM for Composites, an Energy Frontier Research Center funded by the U.S. Department of Energy, Office of Science, Basic Energy Sciences at Clemson University under award # DE-SC0023389.

## Competing interests

The authors declare no competing interests.

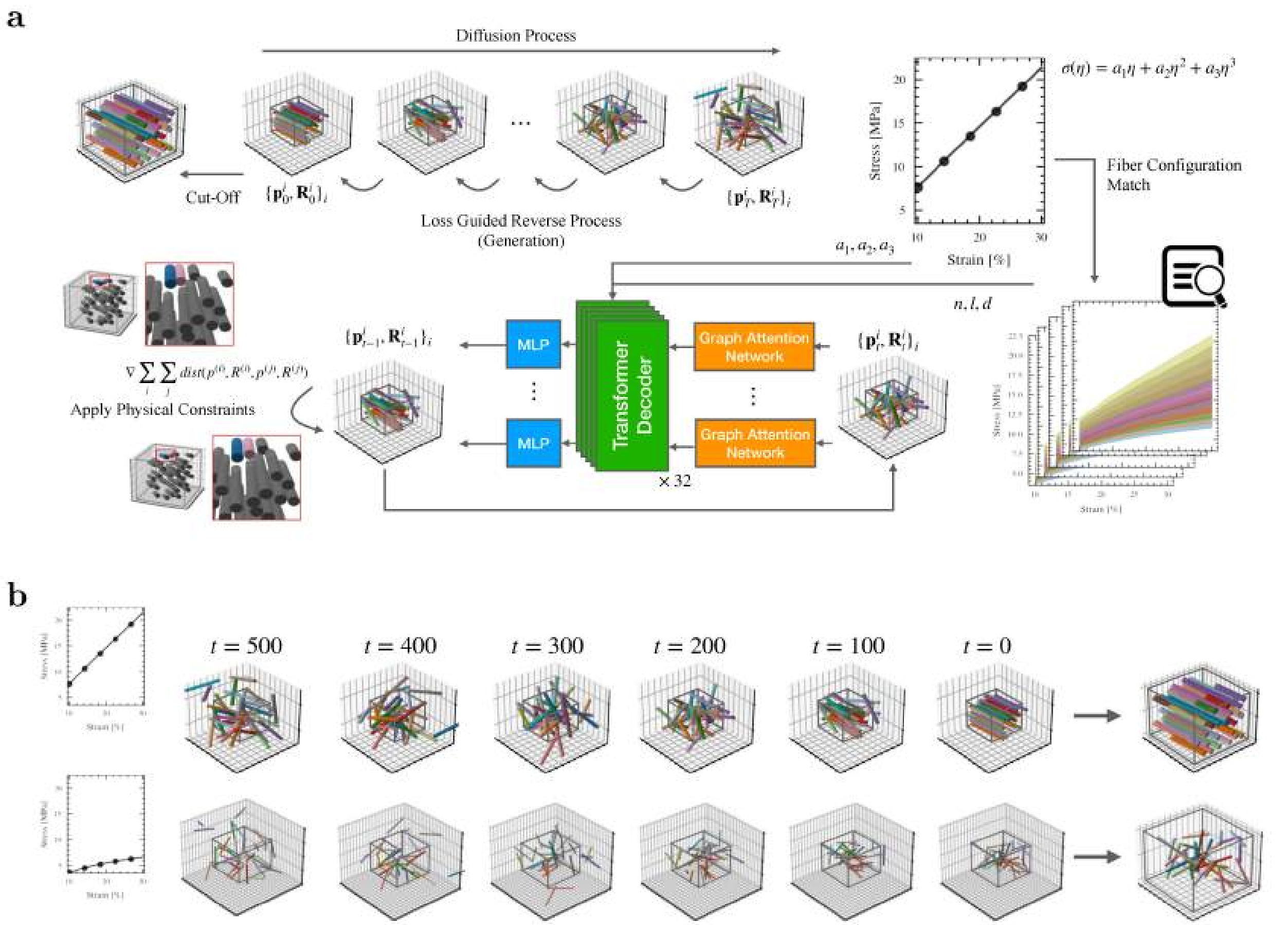

**Figure 1. (a)** Overview of our proposed inverse design system of fiber-reinforced composites. During training, we take the diffusion process and add noise to fiber positions $\boldsymbol{p}_0^i$ and orientations $\boldsymbol{R}_0^i$ by randomly moving and rotating the fibers. A network is trained to predict the noise added to get $\boldsymbol{p}_t^i$ from $\boldsymbol{p}_0^i$. By utilizing the denoising neural network, during inverse design, we estimate $\boldsymbol{p}_{t-1}^i$, $\boldsymbol{R}_{t-1}^i$ from $\boldsymbol{p}_t^i$, $\boldsymbol{R}_t^i$ step by step starting from a randomly drawn fiber spatial distribution of $\{\boldsymbol{p}_T^i, \boldsymbol{R}_T^i\}_i$. The fiber configuration (amount $n$, length $l$ and diameter $d$) is chosen through an automatic matching mechanism that compares the input (expected) stress-strain curve and our collected dataset of the stress-strain response. We employ a stack of 32 transformer decoders as the backbone architecture for noise estimation, coupled with a graph attention network for fiber feature embedding. Additionally, a loss-based guidance is employed during the generation process to apply physical constraints on the generated fiber distributions and to ensure that the generated fiber distributions are free of collisions. **(b)** Demonstrations of the denoising process where arranged fiber distributions ($t = 0$) are achieved from randomly drawn distributions ($t = T$ where $T = 500$ in our implementation) step by step given the expected stress-strain curves shown in the left plots. Top row: a distribution of unidirectional, long fibers penetrating the cubic representative volume element (RVE) with $n = 30$ and $d = 10$. Bottom row: a distribution of short fibers of $n = 20, l = 50, d = 4$ with heterogeneous orientations. In the final step, all fibers are cut along the RVE cube to achieve the spatial distribution where fibers are all inside the RVE.

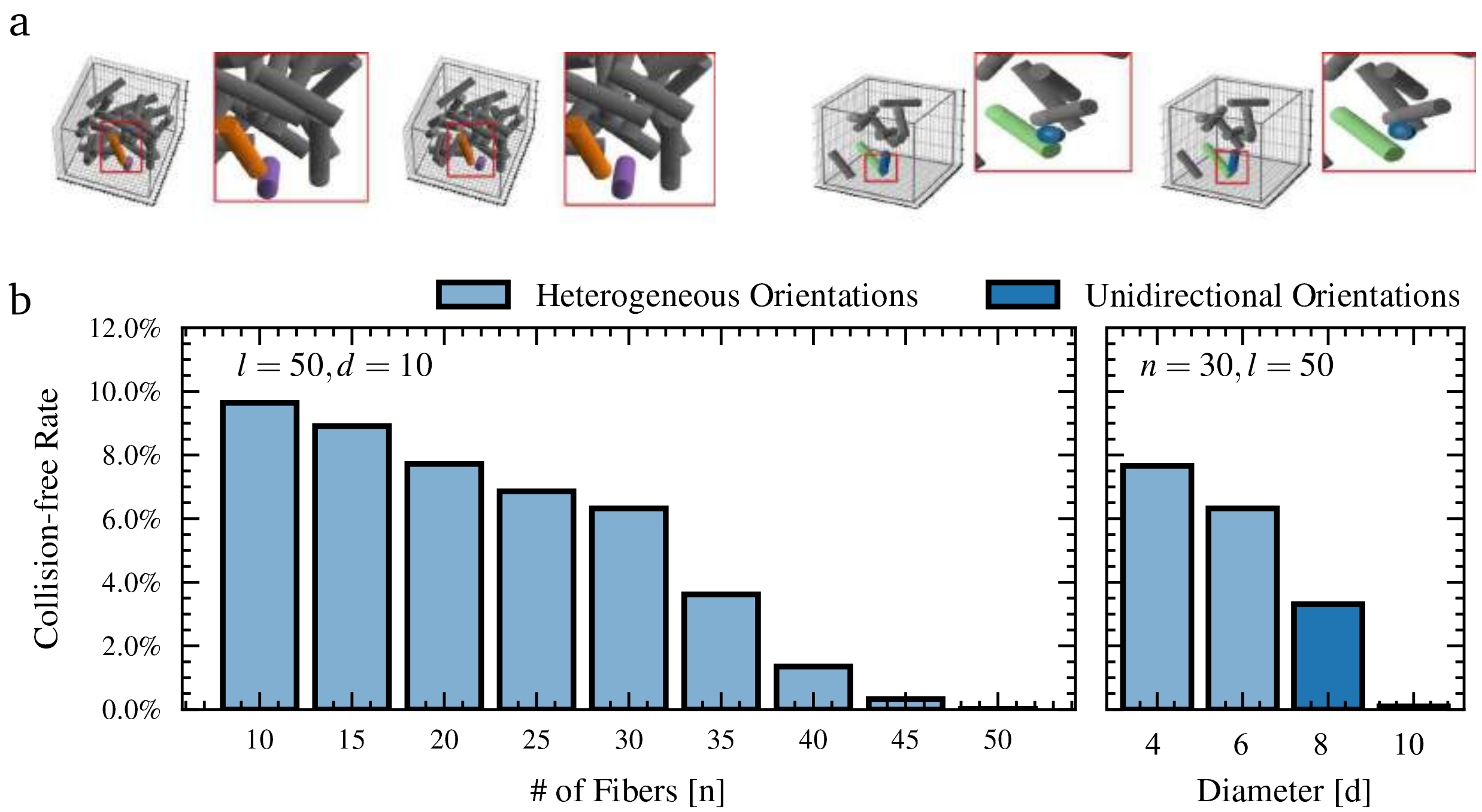

**Figure 2. (a)** Comparison of generated results with (right) and without (left) using the proposed loss guidance. **(b)** Collision-free rate when no loss guidance is employed. The proposed loss guidance can effectively ensure that physical constraints can be applied during the process of generation, and help generate collision-free fiber distributions by slightly modifying the position and/or orientation of the fibers.

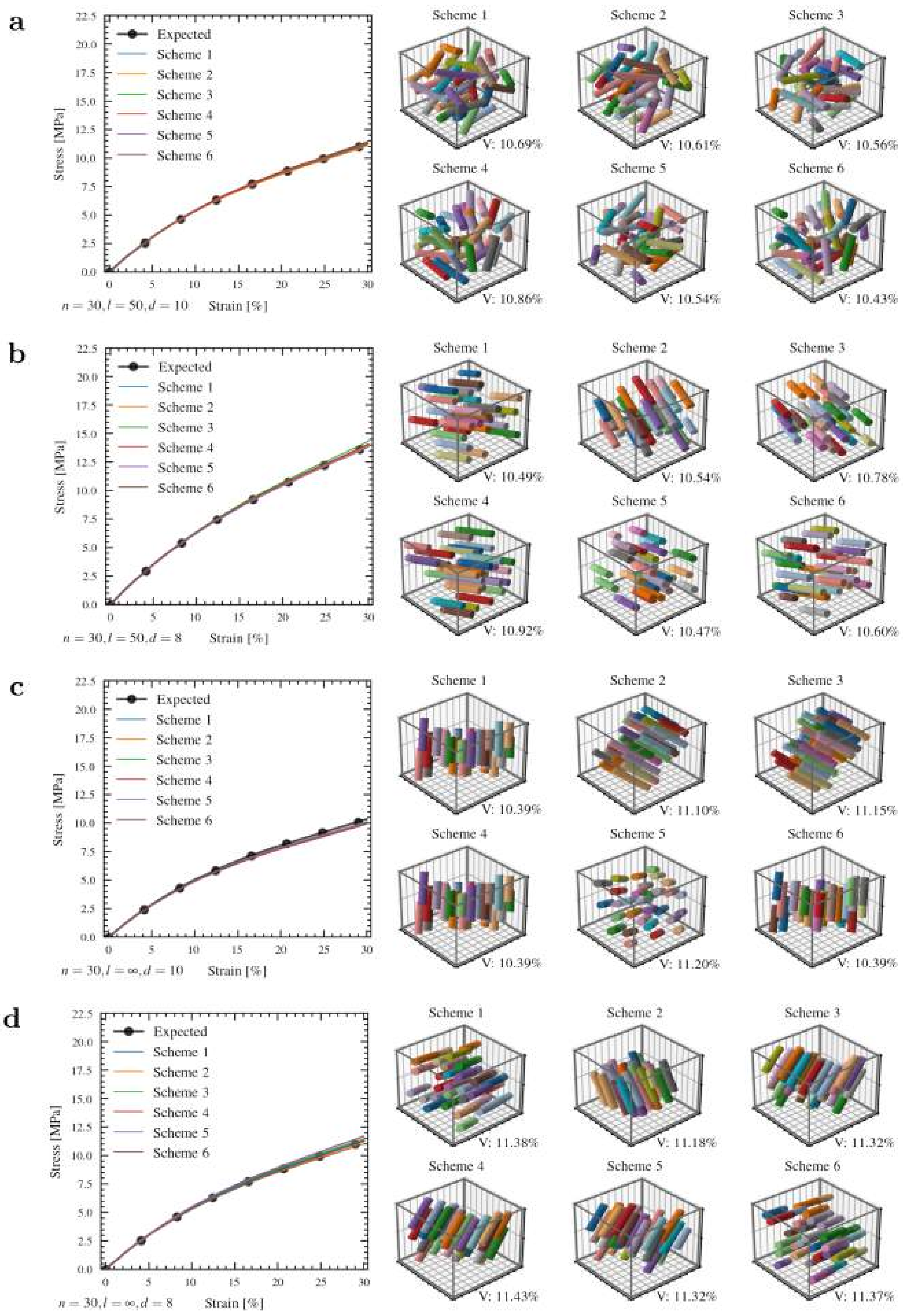

**Figure 3.** Generated fiber distribution schemes using different candidate fiber configurations given the same input stress-strain curve. The four fiber configurations with divergent lengths, diameters, orientation constraints and, thereby, different resulting volume fractions are picked through our automatic matching mechanism. For each candidate fiber configuration, we generate six distinct schemes. Along each scheme, we show the volume fractions ($V$) after cutting off external fiber parts outside the cubic RVE. Our system can find proper fiber configurations and generate various fiber distributions with stress-strain curves close to the given one.

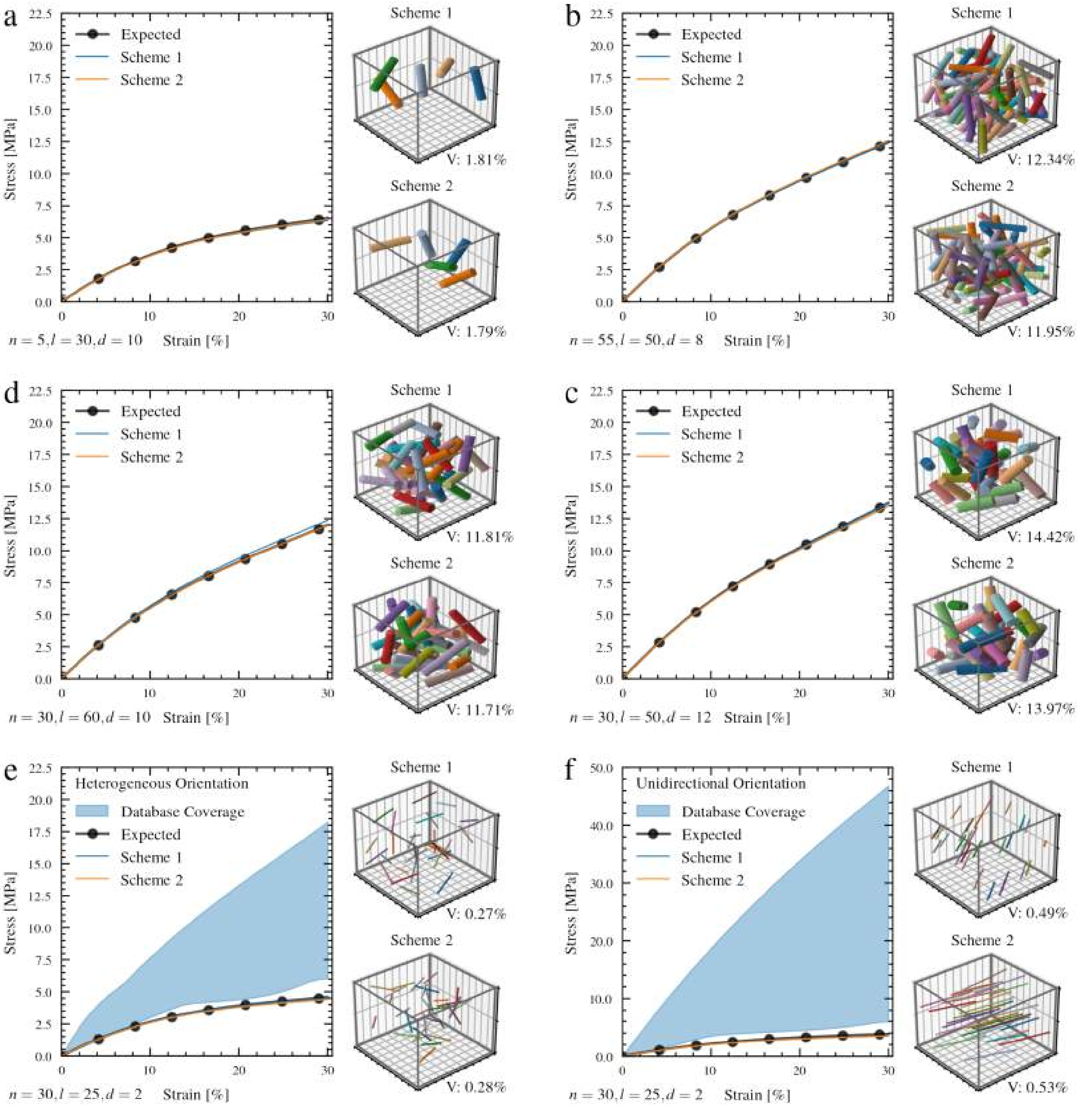

**Figure 4.** Examples of generated results using fiber configurations out of the training set. **(a) - (c):** Results generated using fiber configurations respectively with fiber amount ($n$), length ($l$) and diameter ($d$) not appearing in the training set. **(e) - (f):** Results generated with the target stress-strain curve out of the range covered by the collected database with heterogeneous and unidirectional orientation constraints respectively. This leads to the results where the needed fiber configuration with $l = 25$ and $d = 2$ falls out of the training set.

**Table 1.** Quantitative evaluation using the *best of 10* generated results for each testing sample. Results are reported as relative MAEs in the format of mean $\pm$ std. deviation. For configurations with unidirectional orientations, fiber orientation divergence $e_{ori}$ is reported through the standard deviation in the unit of degrees.

| | $e_{a_1}$ (%) | $e_{a_1}$ (%) | $e_{a_1}$ (%) | $e_A$ (%) | |
|---|---|---|---|---|---|
| *Heterogeneous Orient.* | | | | | |
| $n = 10, l = 50, d = 10$ | 0.04 | 0.05 | 0.07 | 0.04 | |
| $n = 15, l = 50, d = 10$ | 0.03 | 0.04 | 0.05 | 0.03 | |
| $n = 20, l = 50, d = 10$ | 0.06 | 0.15 | 0.19 | 0.06 | |
| $n = 25, l = 50, d = 10$ | 0.05 | 0.06 | 0.06 | 0.05 | |
| $n = 30, l = 50, d = 10$ | 0.05 | 0.06 | 0.12 | 0.06 | |
| $n = 35, l = 50, d = 10$ | 0.05 | 0.09 | 0.19 | 0.05 | |
| $n = 40, l = 50, d = 10$ | 0.09 | 0.11 | 0.24 | 0.09 | |
| $n = 45, l = 50, d = 10$ | 0.07 | 0.17 | 0.47 | 0.07 | |
| $n = 50, l = 50, d = 10$ | 0.05 | 0.07 | 0.14 | 0.05 | |
| $n = 30, l = 50, d = 8$ | 0.03 | 0.04 | 0.07 | 0.03 | |
| $n = 30, l = 50, d = 6$ | 0.04 | 0.10 | 0.19 | 0.04 | |
| $n = 30, l = 50, d = 4$ | 0.09 | 0.04 | 0.02 | 0.09 | |
| $n = 30, l = 30, d = 10$ | 0.02 | 0.03 | 0.03 | 0.02 | |
| $n = 30, l = 30, d = 8$ | 0.04 | 0.06 | 0.07 | 0.03 | |
| $n = 30, l = 30, d = 6$ | 0.02 | 0.05 | 0.07 | 0.02 | |
| $n = 30, l = 30, d = 4$ | 0.04 | 0.02 | 0.03 | 0.04 | |
| *Unidirectional Orient.* | | | | | $e_{ori}$ (%) |
| $n = 30, l = 50, d = 10$ | 0.06 | 0.13 | 0.20 | 0.06 | 0.02 |
| $n = 30, l = 50, d = 8$ | 0.05 | 0.07 | 0.09 | 0.05 | 0.03 |
| $n = 30, l = 50, d = 6$ | 0.04 | 0.11 | 0.17 | 0.04 | 0.02 |
| $n = 30, l = 50, d = 4$ | 0.16 | 0.40 | 0.40 | 0.16 | 0.07 |
| $n = 30, l = \infty, d = 10$ | 0.16 | 0.31 | 0.36 | 0.16 | 0.04 |
| $n = 30, l = \infty, d = 8$ | 0.09 | 0.16 | 0.33 | 0.09 | 0.04 |
| $n = 30, l = \infty, d = 6$ | 0.12 | 0.20 | 0.32 | 0.11 | 0.04 |
| $n = 30, l = \infty, d = 4$ | 0.16 | 0.11 | 0.27 | 0.16 | 0.04 |

**Supplementary Information** for

# Physically Constrained 3D Diffusion for Inverse Design of Fiber-reinforced Polymer Composite Materials

Pei Xu[1,#], Yunpeng Wu[2,#], Srikanth Pilla[3], Gang Li[2,*], and Feng Luo[1,*]

[1]School of Computing, AIM for Composites DOE-Energy Frontier Research Center, Clemson University, Clemson SC 29634

[2]Department of Mechanical Engineering, AIM for Composites DOE-Energy Frontier Research Center, Clemson University, Clemson SC, 29634

[3]Center for Composite Materials, AIM for Composites DOE-Energy Frontier Research Center, Department of Mechanical Engineering, University of Delaware, Newark DE, 19716

*To whom correspondence should be addressed:

Gang Li. Email: gli@clemson.edu

Feng Luo. Email: luofeng@clemson.edu

#These authors contributed equally to the manuscript as first authors.

**Supplementary Note 1: FRPC Data Generation**

Given a specific configuration of fiber length $l$ and diameter $d$ in the unit of mm and amount n, fibers with random positions $\{\boldsymbol{p}^i\}_i$ and orientations $\{\boldsymbol{R}^i\}_i$ are generated inside a cubic representative volume element (RVE) measuring 100mm × 100mm ×100mm. Each fiber is modeled as a homogeneous cylinder. To replicate the experimental tensile test conditions, during the generation process, fibers must reside inside the cubic RVE without any interaction or overlapping with others, which is referred to as collision-free in this work. Any part of the fiber extending beyond the RVE's boundaries is truncated by the RVE's faces.

We generated data of 52 combinations of fiber configurations:

- $n \in [10, 11, \cdots, 49, 50], l = 50, d = 10$ with various orientations,
- $n = 30, l = 30, d \in [4, 6, 8, 10]$ with heterogeneous orientations,
- $n = 30, l = 50, d \in [4, 6, 8, 10]$ with heterogeneous and unidirectional orientations,
- $n = 30, l = \infty, d \in [4, 6, 8, 10]$ with unidirectional orientations.

The volume fractions of the generated FRPCs vary from around 1% to 25%. For the configurations with unidirectional orientations, a unified orientation is randomly sampled and applied to all fibers in one sample, while with heterogeneous orientations each fiber has its orientation sampled randomly and independently. For the scheme with $l = \infty$, fibers are considered long enough to completely penetrate the cubic RVE. To do so, we set $l = 230$ during data generation. For each configuration, we generated 26,000 samples, 25,000 of which are used as the training data and the remaining 1,000 samples are kept for testing. Totally, we have 1.3 million samples for training. We refer to Algorithm 1 for the details of our fiber distribution generation algorithm, where we denote the center position of each fiber using $\boldsymbol{p}^i$ and the orientation using the rotation representation $\boldsymbol{R}^i$. Besides, we assume that the default orientation of fibers with a zero rotation is the positive direction of z-axis, i.e. $\boldsymbol{z}_{ref} := [0, 0, 1]^T$ .

After generating composite microstructures, we employ Gmsh [1] to construct a progressively refined tetrahedron mesh for each generated RVE. Specifically, the top and bottom faces of a fiber cylinder are discretized into identical octagons with 8 vertices for meshing. This leads to an octagonal prism representation of each fiber after discretization and meshing. This technique is exemplified with both the composite RVE and its associated boundary conditions illustrated in Fig. S1.

Once the meshes of a 3D FRPC RVE are produced, the mechanical properties are obtained by using finite element analysis (FEA) with applied displacements of 10mm, 20mm and 30mm, which corresponds to 10%, 20% and 30% strains, respectively. We consider the fibers as linear materials and the matrix of composites as hyperelastic materials for large elastic deformation. The polymer matrix material is described by the 3-parameter Ogden hyperelastic model. The strain energy is given by

$$W = \sum_{i=1}^{3} \frac{\mu_i}{\alpha_i} (\bar{\lambda}_1^{\alpha_i} + \bar{\lambda}_2^{\alpha_i} + \bar{\lambda}_3^{\alpha_i} - 3) + \sum_{k=1}^{3} \frac{1}{D_k} (\mathcal{J} - 1)^{2k}$$

where $\mu_i$ and $\alpha_i$ are material constants in the unit of pressure and a dimensionless quantity respectively, $D_k$ indicates volume change with the initial bulk modulus $K = 2/D_1$, $\lambda_1$, $\lambda_2$ and $\lambda_3$ are principal stretches in three directions of the strain tensor, and $\mathcal{J} = (\lambda_1\lambda_2\lambda_3)^{1/2}$. Numeric values of the material properties used for simulation are listed in Table S1.

Simulation is done on machines equipped with a 16-core E5-2665 CPU and 62GB memory. The time needed to generate and simulate one sample is around 15 seconds to 1 minute mainly depending on the number of fibers.

**Algorithm 1.** Random Fiber Distribution Generation for Data Collection

**function** $Generate\ (\boldsymbol{p}^i, \boldsymbol{R}^i; d, l, n)$
  $a = 0; b = 100$ ▷Boundary constraints
  $r = 0.5d$ ▷Fiber radius
  $i = 0$
  **while** $i < n$ do
    **do**
      $\boldsymbol{p}_i \sim Uniform\ [a + \sqrt{2}r, b - \sqrt{2}r]$ ▷ Fiber position
      Draw $\boldsymbol{R}_i$ in 3D space uniformly. ▷ Fiber orientation
      $\mathrm{p}_i \leftarrow SHRINK(\boldsymbol{p}_i, \boldsymbol{R}_i; r, l, a, b)$ ▷ Apply boundary constraint
      $d_{\min} = \min\{DIST(\boldsymbol{p}^i, \boldsymbol{R}^i; \boldsymbol{p}^j, \boldsymbol{R}^j; d, l)\}_{j<i}$ ▷ Collision Check
    **while** $d_{\min} < d + 0.02$
    $i \leftarrow i + 1$
  **end while**
**end function**

**function** $DIST(\boldsymbol{p}^i, \boldsymbol{R}^i; \boldsymbol{p}^j, \boldsymbol{R}^j; d, l)$ ▷ Shortest distance between two fibers
  $\boldsymbol{d}^i = \boldsymbol{R}^i \boldsymbol{z}_{ref}$ ▷Direction vector of fiber $i$
  $\boldsymbol{u} = l\boldsymbol{d}^i; \boldsymbol{e}^i = \boldsymbol{p}^i - 0.5\boldsymbol{u}$ ▷End points of fiber $i$
  $\boldsymbol{d}^j = \boldsymbol{R}^j \boldsymbol{z}_{ref}; \boldsymbol{v} = l\boldsymbol{d}^j; \boldsymbol{e}^j = \boldsymbol{p}^j - 0.5\boldsymbol{v}$
  $\boldsymbol{r} = \boldsymbol{e}^j - \boldsymbol{e}^i$
  $\Delta = \| \boldsymbol{u} \|^2 \| \boldsymbol{v} \|^2 - \| \boldsymbol{uv} \|^2$
  **if** $\Delta < 10^{-6}$ then
    $s = \boldsymbol{ru} / \| \boldsymbol{u} \|^2; t = s(\boldsymbol{uv} - \boldsymbol{rv}) / \| \boldsymbol{u} \|^2$
  **else**
    $s = \left(Clip\left(\left((\boldsymbol{ru})(\boldsymbol{uv}) - (\boldsymbol{rv})\|\boldsymbol{u}\|^2\right) / \Delta, 0,1\right)(\boldsymbol{uv}) + \boldsymbol{ru}\right) / l^2$
    $t = \left(Clip\left(\left((\boldsymbol{ru})\|\boldsymbol{v}\|^2 - (\boldsymbol{rv})(\boldsymbol{uv})\right) / \Delta, 0,1\right)(\boldsymbol{uv}) - \boldsymbol{rv}\right) / l^2$
  **end if**
  **return** $\|(\mathbf{e}^i + s\mathbf{u}) - (\mathbf{e}^j + t\mathbf{v})\|$
**end function**

**function** $SHRINK(\boldsymbol{p}^i, \boldsymbol{R}^i; r, l, a, b)$ ▷ Shrink a fiber into bounded range
  $\boldsymbol{d}^i = \boldsymbol{R}^i \boldsymbol{z}_{ref}$ ▷ Direction vector of fiber $i$
  $x, y, z \leftarrow \boldsymbol{d}^i$
  $\boldsymbol{d}_a = \left(\left[r\sqrt{x^2 + y^2} + a, r\sqrt{y^2 + z^2} + a, r\sqrt{z^2 + x^2} + a\right]^T - \boldsymbol{p}^i\right) / \boldsymbol{d}^i$
  ▷ Distance to the boundary plane along each axis
  $\boldsymbol{d}_b = \left(\left[b - r\sqrt{x^2 + y^2}, b - r\sqrt{y^2 + z^2}, b - r\sqrt{z^2 + x^2}\right]^T - \boldsymbol{p}^i\right) / \boldsymbol{d}^i$
  **return** $\boldsymbol{p}^i + 0.5\left(\max\{-0.5l, \min\{\boldsymbol{d}_a\}\} + \min\{0.5l, \min\{\boldsymbol{d}_b\}\}\right)\boldsymbol{d}^i$
**end function**

## Supplementary Note 2: Model Architecture

Our network for noise estimation is composed of two modules (Fig. S3): the fiber representation module employs a graph attention network to embed each fiber while taking into account the spatial relationship between the fiber and its neighbors; and the backbone module is a decoder-only transformer architecture taking the embedded representation of fibers as input and outputting the prediction of noise added to fiber positions and orientations.

## SN2.1 Fiber Representation

To enhance the spatial representation of fibers within a RVE, we perform representation learning for each fiber considering the local states of neighbor fibers. For simplicity, we ignore the subscript $t$ here. Then, the $j$-th fiber's state observed by the $i$-th fiber locally is defined as

$$\boldsymbol{n}^{j|i} := [\boldsymbol{p}^{j|i}, \|\boldsymbol{p}^{j|i}\|, \cos(\boldsymbol{d}^j, \boldsymbol{d}^i)]$$

where $j \neq i, \boldsymbol{p}^{j|i} = (\boldsymbol{R}^i)^T(\mathbf{p}^j - \mathbf{p}^i) \in \mathbb{R}^3$ is the relative position of the $j$-th fiber in the $i$-th fiber's local system; $\|\boldsymbol{p}^{j|i}\|$ is the Euclidean distance between fiber $j$ and $i$, $\cos(\boldsymbol{d}^j, \boldsymbol{d}^i) = (\boldsymbol{d}^j)^T \boldsymbol{d}^i$ is the angle between two fibers' orientation directions given the directional vector $\boldsymbol{d}^i = \boldsymbol{R}^i \boldsymbol{z}_{ref}$ and $\boldsymbol{d}^j = \boldsymbol{R}^j \boldsymbol{z}_{ref}$. This leads to $\boldsymbol{n}^{j|i} \in \mathbb{R}^5$ as the neighbor feature of each fiber $j$ related to fiber $i$.

We employ a graph attention network (GAT) [32] to synthesize neighbors. To do so, we regard each fiber as a node in a directional graph linked with edges having weights $e^{j|i}$. The neighbor states then can be synthesized by weights:

$$w^{j|i} = \frac{\exp(e^{j|i})}{\sum_{k \neq i} \exp(e^{j|i})}.$$

The edge weights are obtained through an attention mechanism using the cosine similarity between the self state of the fiber $i$, i.e., $\boldsymbol{p}^i$ and $\boldsymbol{R}^i$, and the local neighbor state $\boldsymbol{n}^{j|i}$:

$$e^{j|i} = \text{LeakyReLU}\left(f_q(\boldsymbol{p}^i, \boldsymbol{d}^i) \cdot f_k(\boldsymbol{n}^{j|i})\right)$$

where the negative slope of LeakyReLU is 0.2. The synthesized neighbor state is obtained by:

$$\sum_j w^{j|i} f_v(\boldsymbol{p}^i, \boldsymbol{d}^i)$$

where $f_q, f_k$ *and* $f_v$ are three embedding neural networks. The corresponding attention mechanism can be written as

$$n^i = \text{Attn}(q, k, v) = \text{Softmax}(f_q(\boldsymbol{p}^i, \boldsymbol{d}^i) \cdot f_k(\{\boldsymbol{n}^{j|i}\}_j)^T \cdot f_v(\{\boldsymbol{p}^i, \boldsymbol{d}^i\}_j)$$

where $(\boldsymbol{p}^i, \boldsymbol{d}^i)$, $\{\boldsymbol{n}^{j|i}\}_j$ *and* $\{\boldsymbol{p}^i, \boldsymbol{d}^i\}_j$ correspond to the query, key and values vectors, respectively, in the vanilla self-attention mechanism. The $i$-th fiber's spatial representation after neighborhood state synthesis is finally denoted as:

$$\boldsymbol{s}^i := [\boldsymbol{p}^i, \boldsymbol{d}^i, \boldsymbol{n}^i].$$

## SN2.2 Backbone Architecture

The backbone architecture of our noise estimation neural network $\epsilon_\theta$ is a transformer [2]. Instead of the encoder-decoder architecture in the vanilla transformer, we employ a decoder-only architecture [3]. Such a structure has been widely used in large language models [4-6] and been demonstrated as an effective architecture to capture the complex relationship between multiple parallel inputs (word tokens in language). In large language models, the neural network is employed to predict the token for the next words recurrently. In our implementation, we utilize the decoder-only architecture to perform noise estimation given multiple inputs (the fiber representations $s^i$) parallelly at the same time, while the condition $c$, including fiber diameter $d$ and length $l$, and the time step $t$ are fed into the network as the memory input. Since the transformer architecture adopts attention mechanisms and provides an output having the same dimension as the input, we do not need to explicitly indicate the fiber amount $n$ in the condition vector.

Another advantage of transformer architecture is that it is an architecture invariant to the order of input. When applying to natural language processing tasks, the transformer gets the word order information through positional encodings added to each input word token. However, our task is input permutation invariant. Namely, the input order of fibers should not impact the output of the network for each fiber. This characteristic is naturally supported by the transformer architecture without using positional encoding. Instead, we apply the position encoding technique to encode the time step variable $t$. Following the previous notation $\bar{t} = t/T$, we define the embedded time step $t$ through sinusoidal positional encoding. This leads to a fiber configuration vector $c = [d, l]$ and a timestep vector $t = [\sin\bar{t}, \cos\bar{t}]$ serving together as the condition input to the transformer network. The output of the transformer corresponding to each fiber representation $\boldsymbol{s}_t^i$ is finally passed through a shared two-layer neural network to get the predicted noise $[\boldsymbol{p}_{\epsilon_t}^i, \omega_{\epsilon_t}^i \boldsymbol{u}_{\epsilon_t}^i] \in \mathbb{R}^5$ . The architecture hyperparameters of the backbone transformer are listed in Table S2.

**Supplementary Note 3: Training Protocol**

Our condition vector $c$ does not include the configuration of unidirectional orientation or not. We train two models to handle the configuration with and without unidirectional orientations respectively. The model is trained using the loss function defined by Equations 8, 9 and 10. We list the hyperparameters that we used during model training and diffusion process in Table S3. For the rotation noise variance $\sigma_t$, when $\sigma_1 = 5$, the probability density function $f_{IGSO}$ (Equation 6) will lead to a linearly increasing cumulative distribution function, and results in a distribution where the angle $\omega$ is distributed almost uniformly within the range of $[0, \pi]$. Because of the cyclical nature of angles in the range of $[-\pi, \pi]$, variance exploding form of diffusion process then will not lead to any basis caused by the ground truth original value. Therefore, in the reverse process, we can draw $\boldsymbol{R}_T^i$ through $IGSO$ directly with an random angle distributed in the range of $[0, \pi]$ uniformly, and then scale it by $1/2$ to get a rotation in the range of $[0, \pi/2]$ as the initial value for data generation, given the homogeneous nature of fibers. All our models are trained on four machines, each of which is equipped with 2×A100 Nvidia GPUs. The whole training consuming all the training data takes around 1 week.

**Supplementary Note 4: Inference Protocol for Inverse Design**

During inference, our model takes a set of user-provided stress-strain curve coefficients $(a_1, a_2, a_3)$ as input and outputs fiber distributions $\{\boldsymbol{p}^i, \boldsymbol{R}^i\}_i$ whose stress-strain responses can meet the provided target curve. After receiving the expected target curve coefficients, an automatic matching module will first propose the potential fiber configurations ($n$, $l$, and $d$) by checking the collected database. A fiber configuration is considered valid if its stress-strain response range can cover the provided target curve. After deciding the fiber configuration, we will have the condition vector $c = [d, l]$. The fiber amount $n$ is implicitly imposed by using a n-element set of fiber distributions $\{\boldsymbol{p}_t^i, \boldsymbol{R}_t^i\}_{i=1}^n$, where $t = 0, \cdots, T$. The initial fiber positions $\{\boldsymbol{p}_T^i\}_{i=1}^n$ are drawn from the standard normal distribution. For fiber orientation proposals $\{\boldsymbol{R}_T^i\}_{i=1}^n$, we can draw the rotation angle $\{\omega_T^i\}_i$ directly from $[0, \pi]$ uniformly and scaled by $1/2$ as the initial value combing with a randomly generated 2D unit vector $\boldsymbol{\mu}_T^i$. Then we perform the reverse process of the diffusion model to iteratively obtain $\{\boldsymbol{p}_{t-1}^i, \boldsymbol{R}_{t-1}^i\}_i$ from $\{\boldsymbol{p}_T^i, \boldsymbol{R}_T^i\}_i$ .

To generate physically plausible results, we introduce loss guidance to optimize the generated results $\{\boldsymbol{p}_{t-1}^i, \boldsymbol{R}_{t-1}^i\}_i$ by running the gradient descent method to minimize Equation 1 once at each timestep t. At the last step from $t = 1$ to $t = 0$, we keep running the gradient descent method to minimize Equation 1 until it converges to zero. Typically, at the last step, the loss function would reach zero within less than 10 iterations. The loss guidance would not change the generated distribution too much, while guaranteeing to generate collision-free fiber distributions by slightly moving or rotating the fibers inside the RVE.

The whole reverse process with loss guidance is described in Algorithm 2.

**Algorithm 2.** Reverse Process for Fiber Inverse Design

Initialize fiber position $\mathbf{p}_T^i \sim \mathcal{N}(\mathbf{0}, \boldsymbol{I}) \in \mathbb{R}^3$ and rotation $\boldsymbol{R}_T^i = \omega_T^i \boldsymbol{\mu}_T^i$, where $\boldsymbol{\omega}_T^i \sim Uniform(0, \pi)$, $\boldsymbol{\mu}_T^i = \tilde{\boldsymbol{\mu}}_T^i / \|\tilde{\boldsymbol{\mu}}_T^i\|$, $\tilde{\boldsymbol{\mu}}_T^i \sim \mathcal{N}(\mathbf{0}, \boldsymbol{I}) \in \mathbb{R}^2$, and $i = 0, \cdots, n$.
**for** $t \leftarrow T$ to 1 **do**
  Get $\{\boldsymbol{p}_{t-1}^i, \boldsymbol{R}_{t-1}^i\}_i$ from $\{\boldsymbol{p}_T^i, \boldsymbol{R}_T^i\}_i$ based on Equation 10, 12.
  **if** $t - 1 = 0$ **then** ▷ Apply physical constrains via loss guidance
    **while** $\mathcal{L}_{\text{cons}}\left(\{\boldsymbol{p}_{t-1}^i, \boldsymbol{R}_{t-1}^i\}_i\right) > 0$ **do**
      Update $\{\boldsymbol{p}_{t-1}^i, \boldsymbol{R}_{t-1}^i\}_i$ based on Equation 1.
    **end while**
  **else**
    Update $\{\boldsymbol{p}_{t-1}^i, \boldsymbol{R}_{t-1}^i\}_i$
  **end if**
**end for**

**Table S1.** Material properties used during simulation of finite element analysis.

| ***Fibers*** | | ***RVE* (3-Parameter Ogden Hyperelastic Model)** | | | | | |
|---|---|---|---|---|---|---|---|
| Elastic Modulus ($E$) | 1000MPa | $a_1$ | 2.74 | $\mu_1$ | -9.19 MPa | $D_1$ | 0.00001 |
| Poisson's Ratio ($v$) | 0.3 | $a_2$ | -5.55 | $\mu_2$ | -8.61 MPa | | |
| | | $a_3$ | 1.31 | $\mu_3$ | -6.92 MPa | | |

**Table S2.** Architecture of the backbone transformer.

| | | | |
|---|---|---|---|
| Number of Attention Heads | 16 | Number of Stacked Decoders | 32 |
| Input Dimension | 512 | Nonlinear Activator | ReLU |
| Latent Dimension of Feedforward Network | 2048 | Total Parameters | 150M |

**Table S3.** Hyperparameters of PC3D_Diffusion.

| ***Model Training*** | | ***Diffusion Process*** | | | | | |
|---|---|---|---|---|---|---|---|
| Learning Rate | 0.0003 | $T$ | 500 | $\sigma_0$ | 0.05 | $\varepsilon$ | 0.002 |
| Optimizer | AdamW | $\beta_1$ | 0.0001 | $\sigma_1$ | 5 | $\alpha$ | 0.001 |
| Batch Size | 256 | $\beta_T$ | 0.02 | $b_0$ | 1 | | |

**Table S4**. Quantitative evaluation averaged over 10 generated results for each testing sample. Results are reported as relative MAEs. For configurations with unidirectional orientations, fiber orientation divergence $e_{ori}$ is reported through the standard deviation in the unit of degrees.

| | $e_{a_1}$ (%) | $e_{a_1}$ (%) | $e_{a_1}$ (%) | $e_A$ (%) | |
|---|---|---|---|---|---|
| *Heterogeneous Orientations* | | | | | |
| $n = 10, l = 50, d = 10$ | 0.21 ± 0.26 | 0.21 ± 0.33 | 0.27 ± 0.51 | 0.21 ± 0.26 | |
| $n = 15, l = 50, d = 10$ | 0.20 ± 0.22 | 0.19 ± 0.18 | 0.24 ± 0.24 | 0.20 ± 0.22 | |
| $n = 20, l = 50, d = 10$ | 0.36 ± 0.26 | 0.78 ± 0.60 | 0.91 ± 0.76 | 0.36 ± 0.26 | |
| $n = 25, l = 50, d = 10$ | 0.28 ± 0.30 | 0.30 ± 0.27 | 0.40 ± 0.38 | 0.28 ± 0.30 | |
| $n = 30, l = 50, d = 10$ | 0.37 ± 0.43 | 0.39 ± 0.41 | 0.55 ± 0.64 | 0.37 ± 0.43 | |
| $n = 35, l = 50, d = 10$ | 0.32 ± 0.33 | 0.45 ± 0.45 | 0.66 ± 0.75 | 0.32 ± 0.33 | |
| $n = 40, l = 50, d = 10$ | 0.36 ± 0.37 | 0.49 ± 0.58 | 0.82 ± 0.78 | 0.36 ± 0.37 | |
| $n = 45, l = 50, d = 10$ | 0.44 ± 0.38 | 0.72 ± 0.86 | 0.93 ± 0.70 | 0.44 ± 0.38 | |
| $n = 50, l = 50, d = 10$ | 0.28 ± 0.20 | 0.44 ± 0.32 | 0.79 ± 0.63 | 0.28 ± 0.20 | |
| $n = 30, l = 50, d = 8$ | 0.18 ± 0.17 | 0.20 ± 0.18 | 0.29 ± 0.30 | 0.18 ± 0.17 | |
| $n = 30, l = 50, d = 6$ | 0.22 ± 0.20 | 0.40 ± 0.50 | 0.53 ± 0.91 | 0.22 ± 0.19 | |
| $n = 30, l = 50, d = 4$ | 0.27 ± 0.15 | 0.17 ± 0.13 | 0.14 ± 0.14 | 0.27 ± 0.15 | |
| $n = 30, l = 30, d = 10$ | 0.13 ± 0.11 | 0.14 ± 0.10 | 0.19 ± 0.19 | 0.13 ± 0.11 | |
| $n = 30, l = 30, d = 8$ | 0.26 ± 0.17 | 0.32 ± 0.25 | 0.40 ± 0.33 | 0.26 ± 0.17 | |
| $n = 30, l = 30, d = 6$ | 0.12 ± 0.09 | 0.28 ± 0.21 | 0.34 ± 0.26 | 0.12 ± 0.09 | |
| $n = 30, l = 30, d = 4$ | 0.14 ± 0.09 | 0.11 ± 0.08 | 0.11 ± 0.15 | 0.15 ± 0.09 | |
| *Unidirectional Orientations* | | | | | $e_{ori}$ (%) |
| $n = 30, l = 50, d = 10$ | 0.38 ± 0.31 | 0.85 ± 0.64 | 1.07 ± 1.04 | 0.38 ± 0.31 | 0.04 |
| $n = 30, l = 50, d = 8$ | 0.28 ± 0.23 | 0.39 ± 0.33 | 0.53 ± 0.47 | 0.28 ± 0.23 | 0.04 |
| $n = 30, l = 50, d = 6$ | 0.20 ± 0.20 | 0.38 ± 0.70 | 0.50 ± 1.19 | 0.20 ± 0.19 | 0.04 |
| $n = 30, l = 50, d = 4$ | 0.73 ± 0.73 | 1.39 ± 1.13 | 1.48 ± 1.00 | 0.74 ± 0.73 | 0.31 |
| $n = 30, l = \infty, d = 10$ | 0.96 ± 0.79 | 1.71 ± 1.46 | 2.42 ± 1.42 | 0.96 ± 0.79 | 0.27 |
| $n = 30, l = \infty, d = 8$ | 0.55 ± 0.48 | 0.97 ± 0.91 | 1.47 ± 1.69 | 0.55 ± 0.48 | 0.12 |
| $n = 30, l = \infty, d = 6$ | 0.55 ± 1.27 | 0.73 ± 1.80 | 0.82 ± 1.46 | 0.54 ± 1.27 | 0.10 |
| $n = 30, l = \infty, d = 4$ | 0.37 ± 0.57 | 0.33 ± 0.65 | 0.65 ± 0.66 | 0.37 ± 0.57 | 0.10 |

**Table S5**. Sensitive analysis of different transformer architectures and the analysis using and not using GAT for fiber spatial representation learning. All results are studied using the configuration of $n = 30, l = 50, d = 10$ and reported using relative MAEs averaged over 10 generated results for each testing sample.

| Heads | Layers | GAT | $e_{a_1}$ [%] | $e_{a_1}$ [%] | $e_{a_1}$ [%] | $e_A$ [%] |
|---|---|---|---|---|---|---|
| 1 | 8 | ✓ | 0.13 | 0.18 | 0.27 | 0.13 |
| 1 | 16 | ✓ | 0.10 | 0.13 | 0.21 | 0.10 |
| 4 | 16 | ✓ | 0.09 | 0.12 | 0.19 | 0.09 |
| 8 | 16 | ✓ | 0.09 | 0.10 | 0.18 | 0.09 |
| 8 | 32 | ✓ | 0.07 | 0.07 | 0.14 | 0.07 |
| 16 | 32 | ✓ | **0.05** | **0.06** | **0.12** | **0.06** |
| 16 | 32 | - | 0.08 | 0.08 | 0.16 | 0.08 |

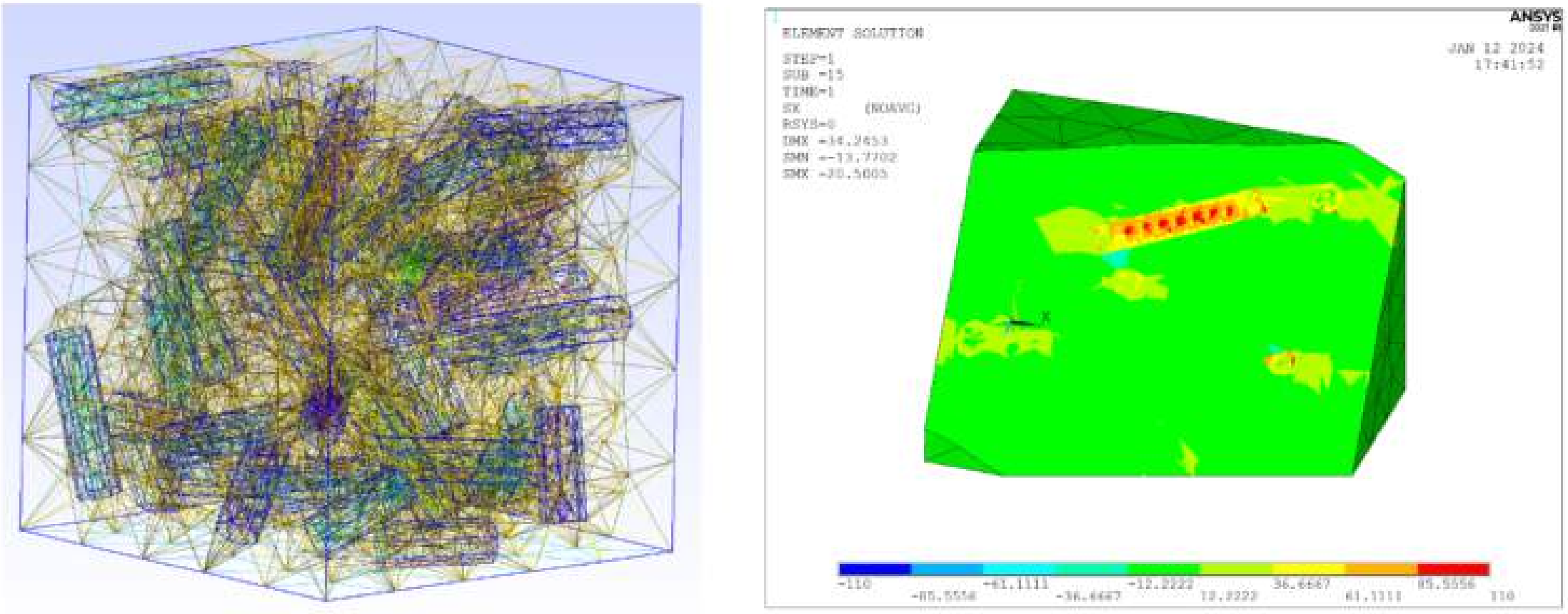

**Fig. S1.** Demonstration of structural analysis on generated data. **Left:** meshing results by discretizing each fiber as an octagonal prism. **Right:** stress response obtained by simulation using finite element analysis.

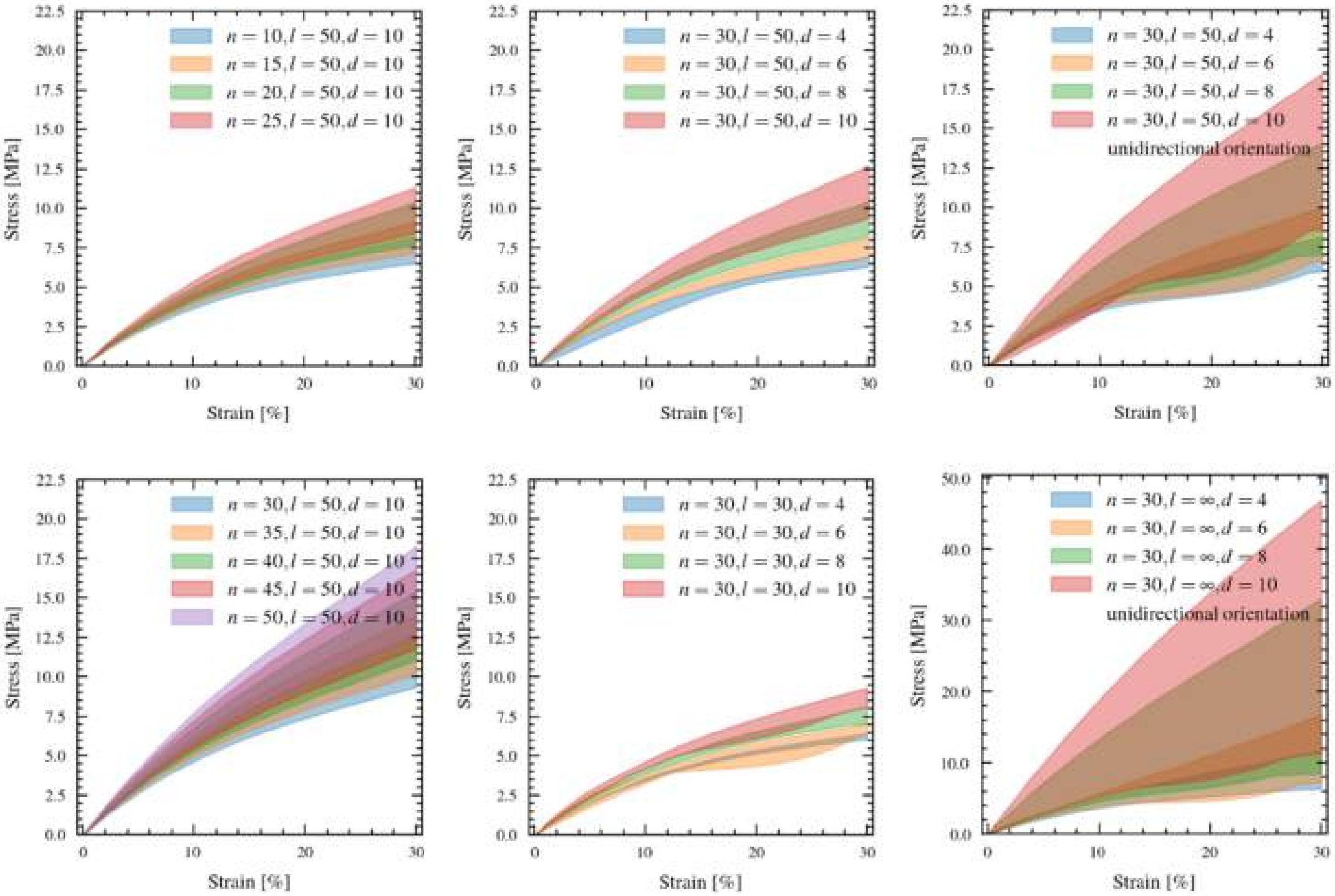

**Fig. S2.** Ranges of the stress-strain curves of different fiber configurations.

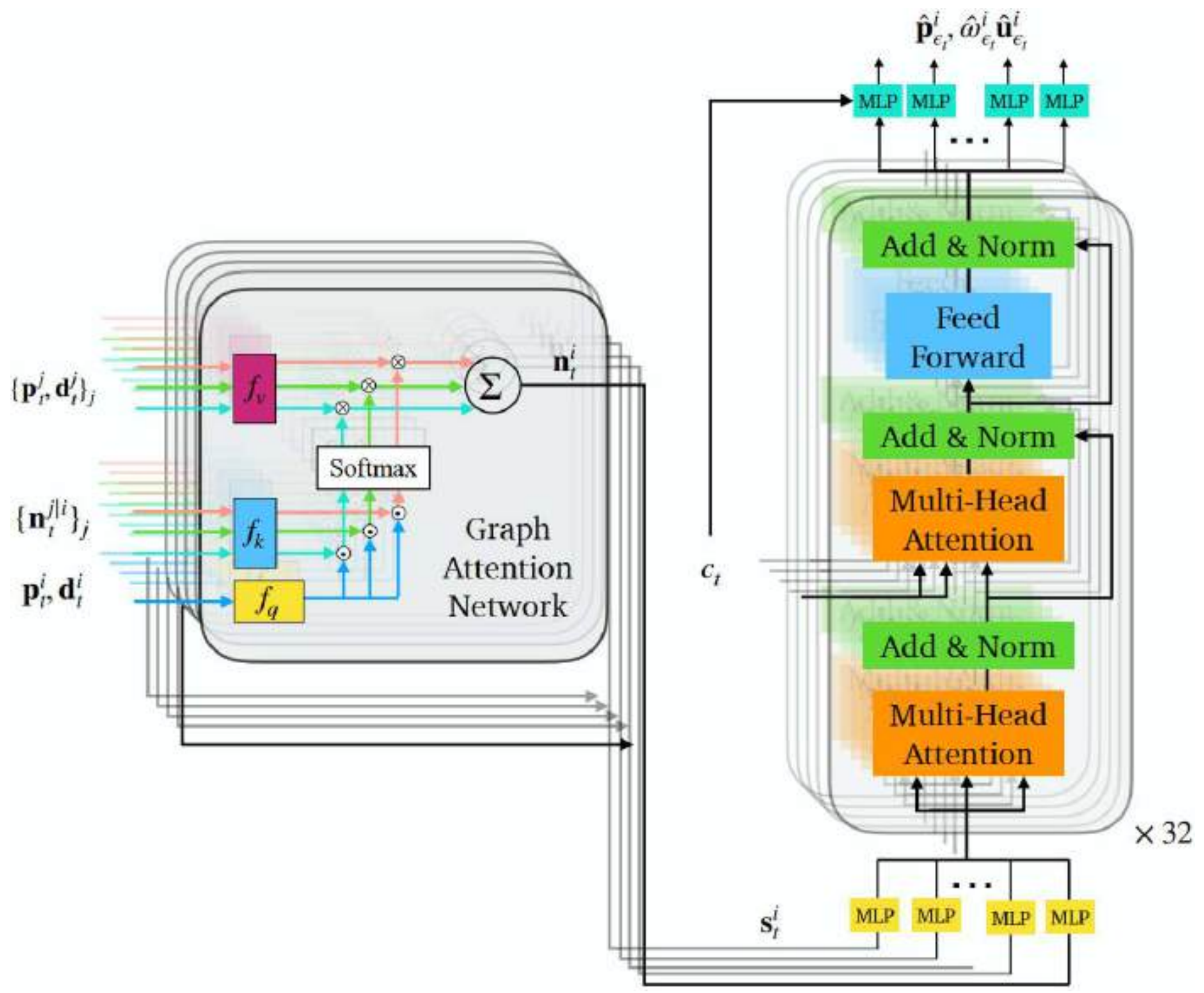

**Fig. S3.** Network architectures $\epsilon_\theta$ for noise estimation. Left: graph attention network to enhance the spatial representation of each fiber $i$ by introducing the states of neighbor fibers $\{\boldsymbol{p}_t^j, \boldsymbol{d}_t^j\}_j$ through their local spatial states $\{\boldsymbol{n}_t^{j|i}\}_j$ relative to the fiber $i$. Right: transformer decoders used to predict added noise given the condition state $c_t$ and fiber representations $s_t^i$. The synthesized fiber state $s_t^i$ is projected to the space of $\mathbb{R}^{512}$ through a multilayer perception (yellow), and the final output of the decoders is converted to the prediction $[\hat{\boldsymbol{p}}_{\epsilon_t}^i, \hat{\omega}_{\epsilon_t}^i \hat{\boldsymbol{u}}_{\epsilon_t}^i] \in \mathbb{R}^5$ corresponding to each input fiber state $s_t^i$ through an additional multilayer perception (cyan).

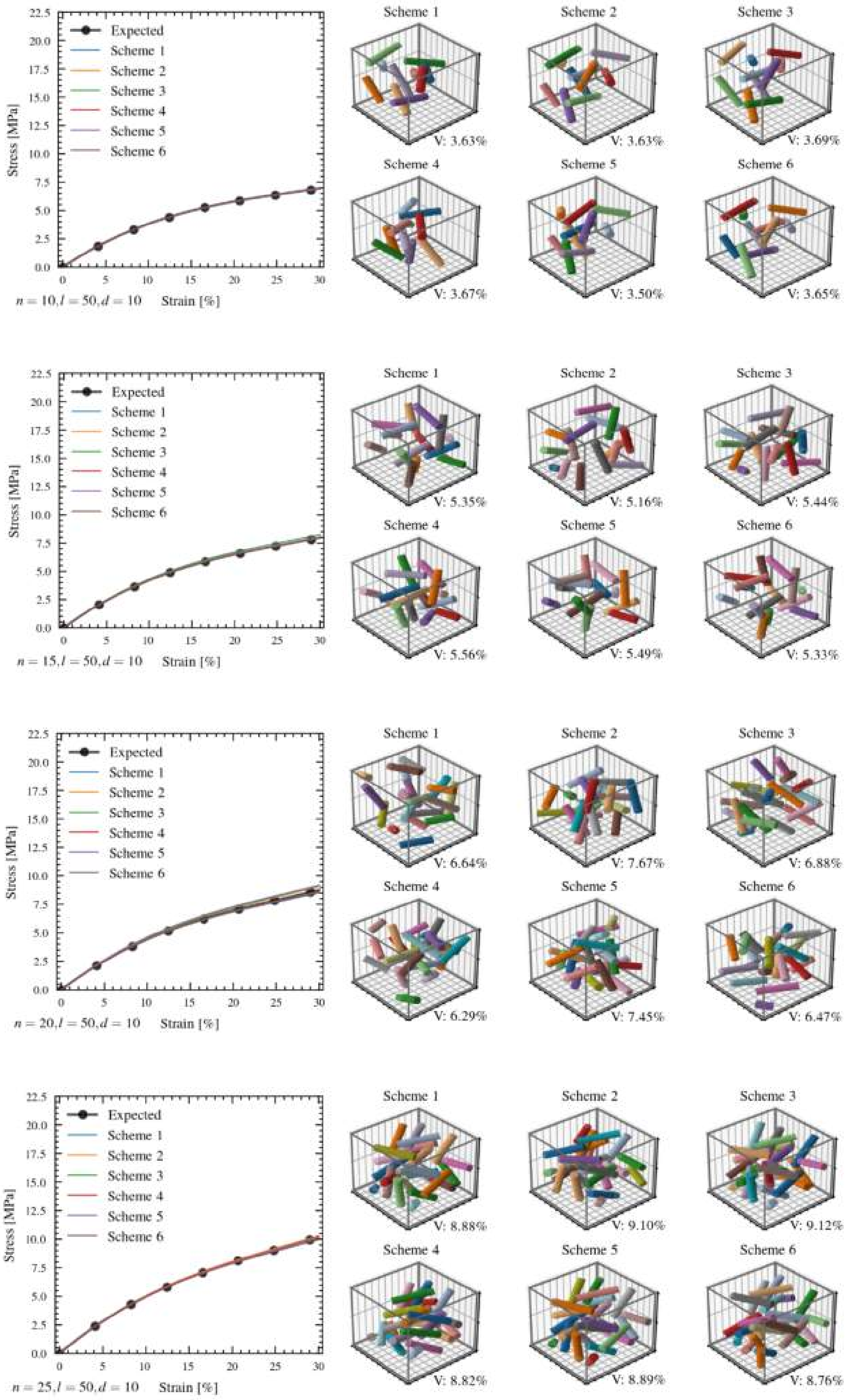

**Fig. S4**. Generated short fiber distribution schemes with different numbers of fibers ($n = 10, 15, 20, 25$). All fibers distributed partially outside the cubic RVE are cut off along each side of the RVE. Volume fractions (V) are computed after cut-off.

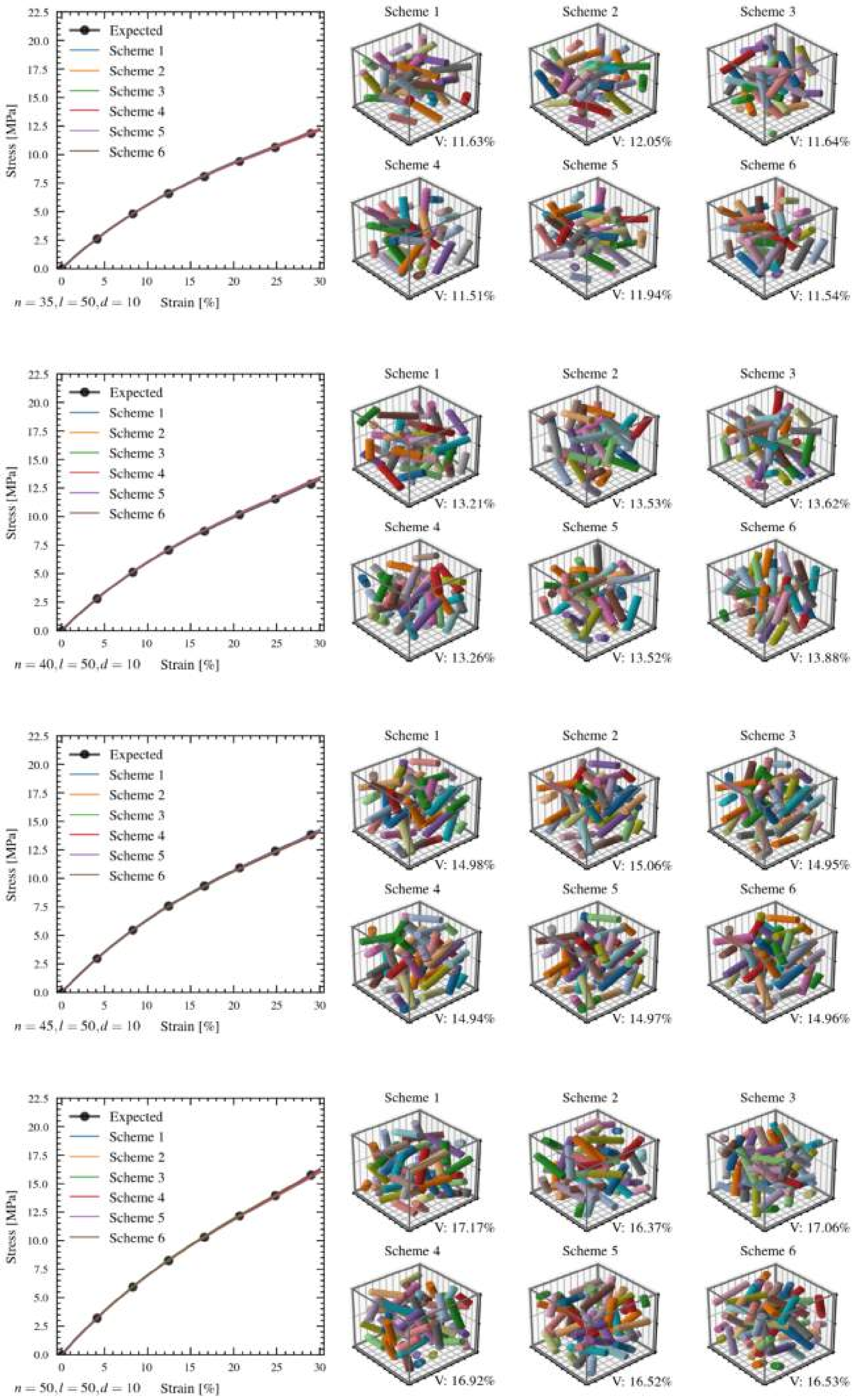

**Fig. S5**. Generated short fiber distribution schemes with different numbers of fibers ($n = 35, 40, 45, 50$).

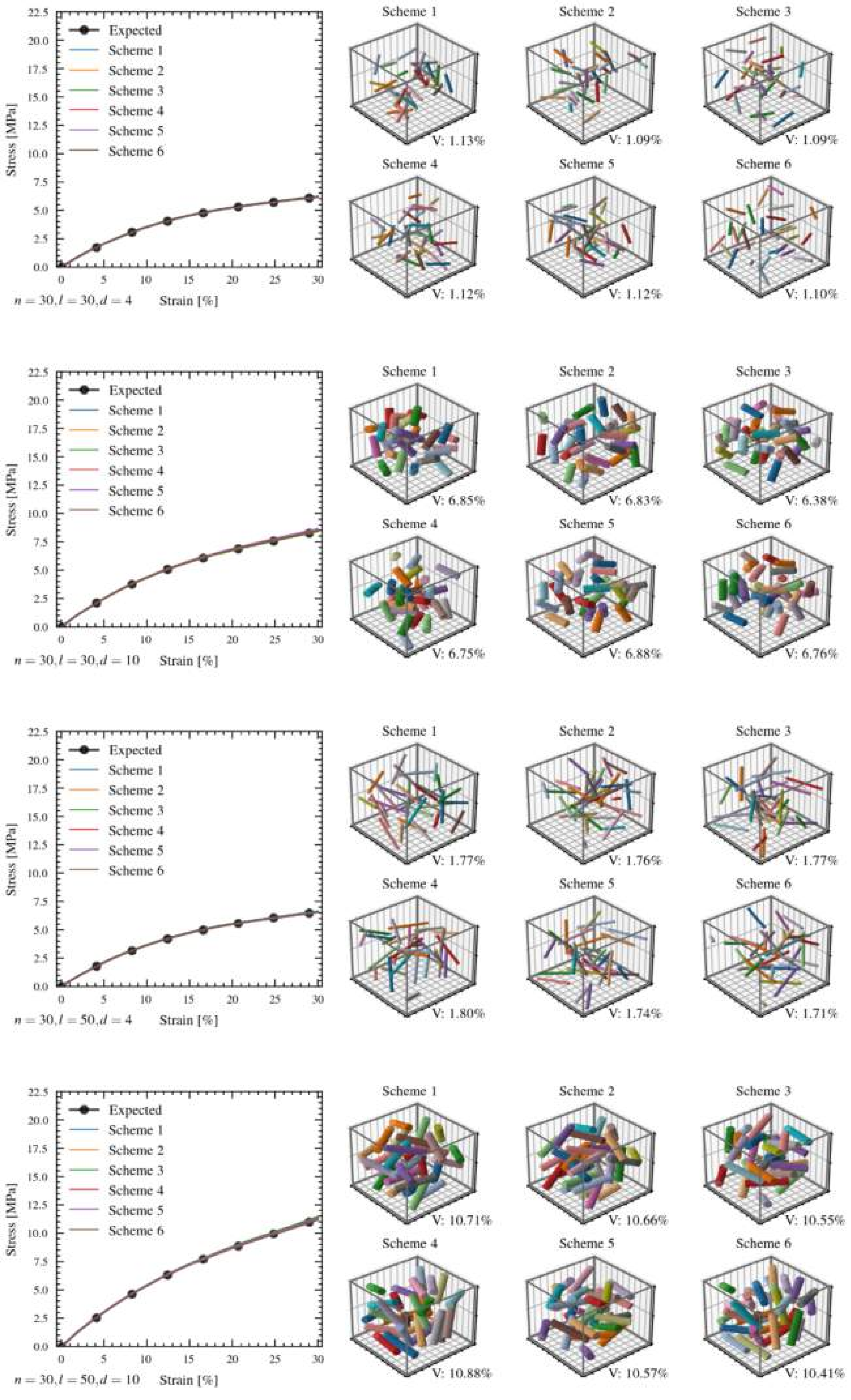

**Fig. S6**. Generated unidirectional, short fiber distribution schemes given $n = 30, l = 50$ with different fiber diameters.

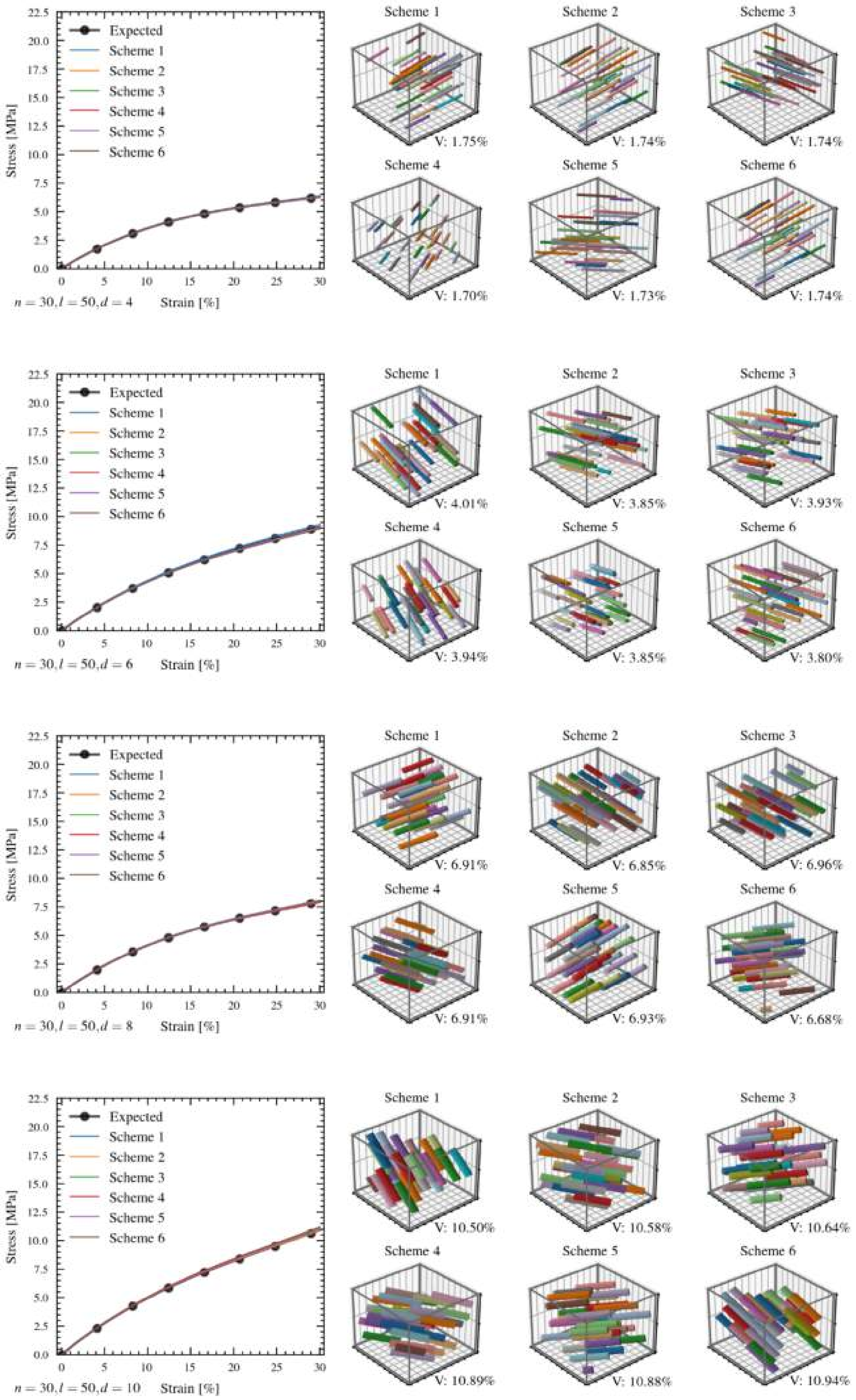

**Fig. S7**. Generated unidirectional, short fiber distribution schemes given $n = 30, l = 50$ with different fiber diameters.

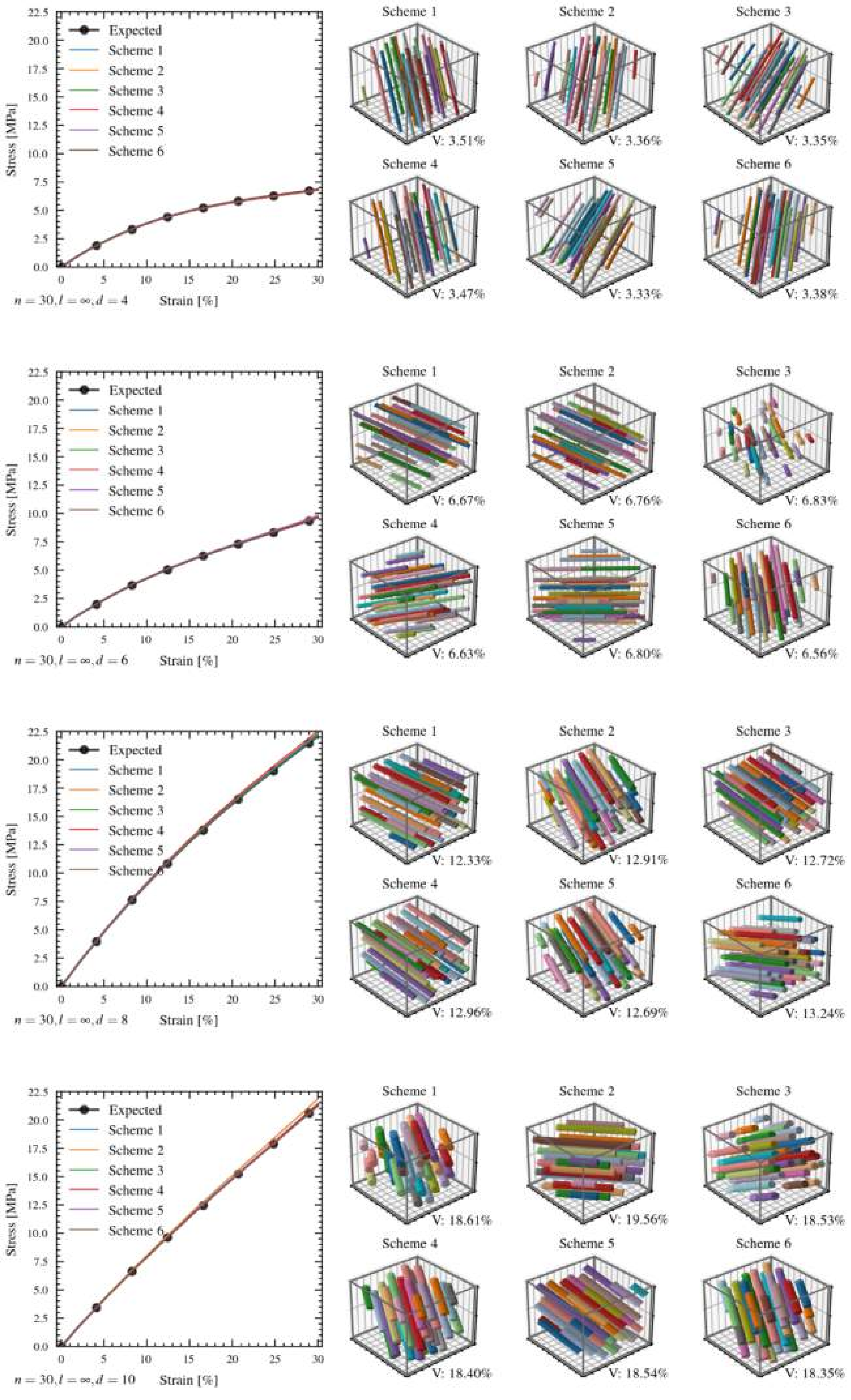

**Fig. S8**. Generated unidirectional, long fiber distribution schemes with different fiber diameters.

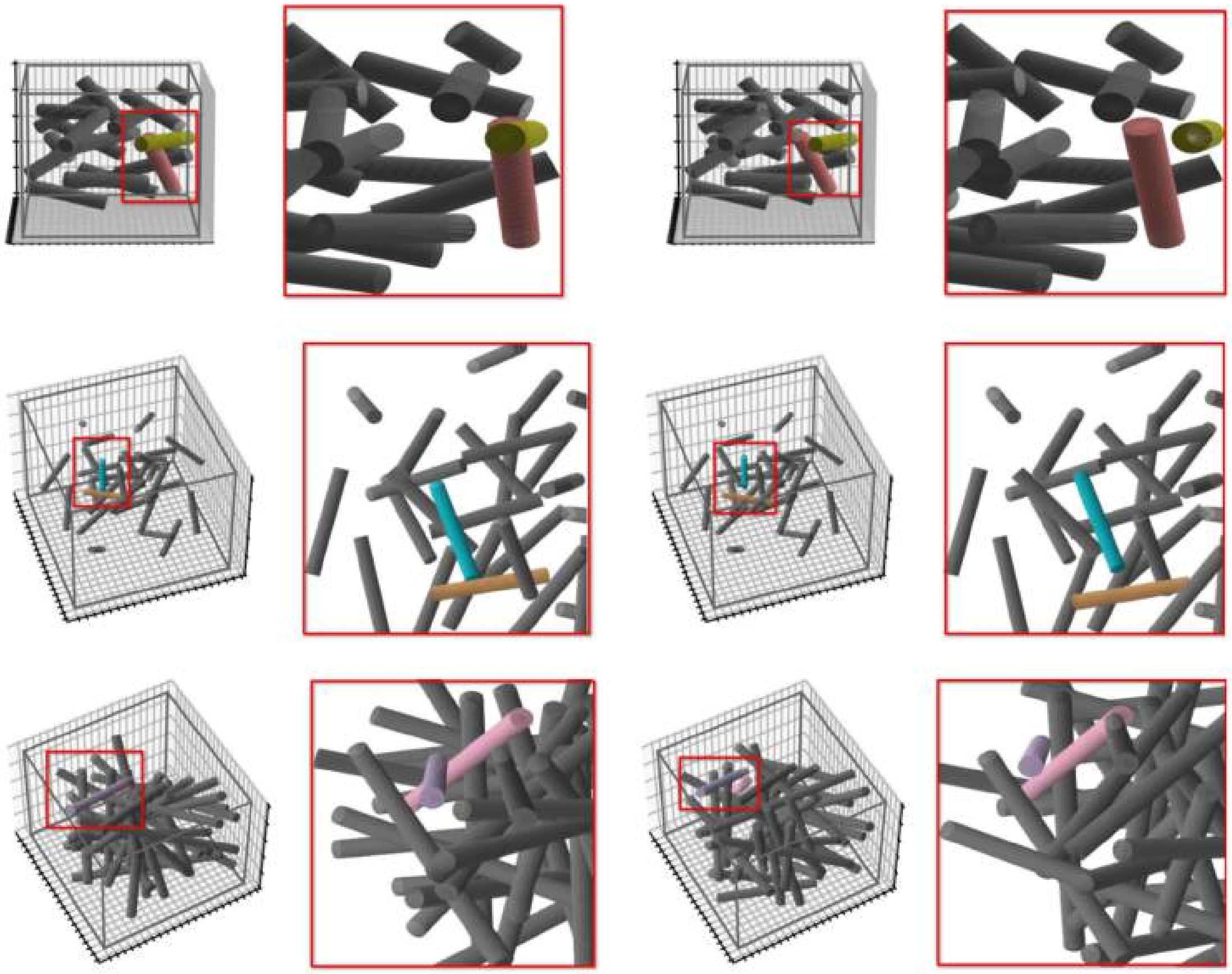

**Fig. S9**. Comparison between the generated results without applying the loss guidance during generation (left) and those using the loss guidance to ensure the physical constraints (right).

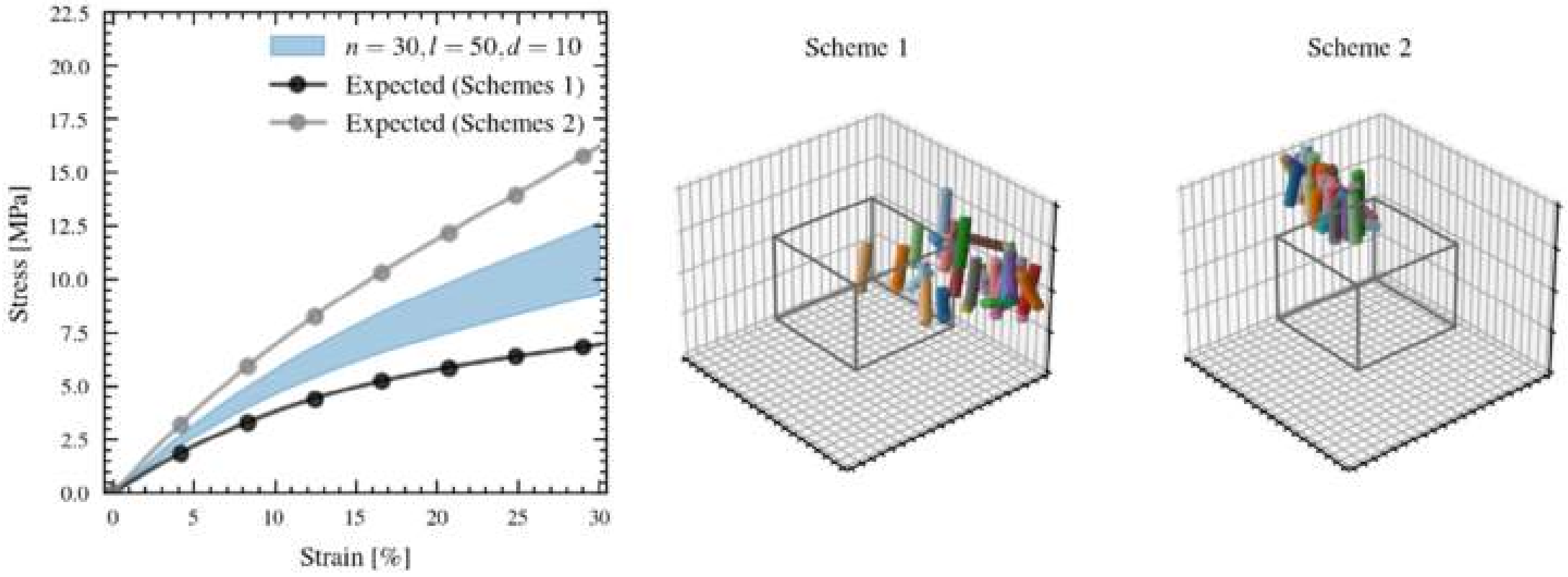

**Fig. S10**. Failure cases where the model is asked to generate fiber distributions with an expected stress-strain curve outside the valid range. The blue range indicates the valid range of the stress-strain curves given the target fiber configuration $n = 30, l = 50, d = 10$. The two marked lines are the input curves fed to the model. The two small plots show the generated results.

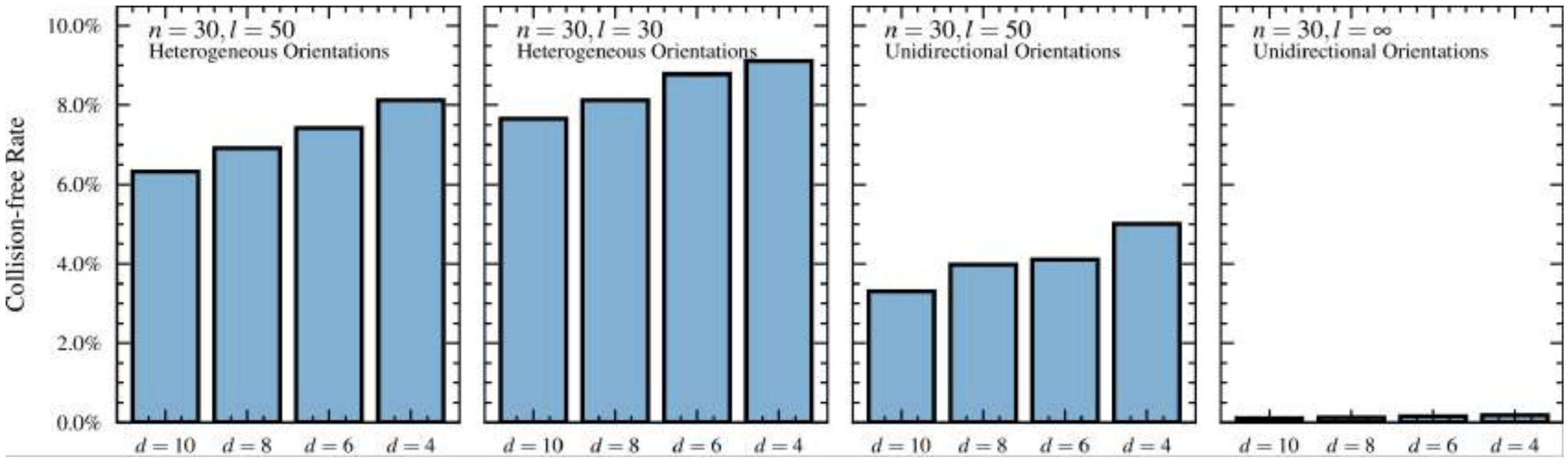

**Fig. S11**. Collision-free rate when generating fiber distributions without using the proposed loss guidance.